\documentclass[a4paper,11pt]{article}
\pdfoutput=1 
\usepackage{jheppub} 
\usepackage{amsmath,amssymb}
\usepackage{datetime,mathdots}

\newcommand\TM{Teichm\"uller~}

\newcommand{\ud}{\mathrm{d}}
\newcommand{\ub}{\mathrm{b}}
\newcommand{\R}{\mathrm{R}}
\newcommand{\NS}{\mathrm{NS}}

\newcommand{\xione}{{\xi_1}}
\newcommand{\xitwo}{{\xi_2}}

\def\be{\begin{equation}}
\def\ee{\end{equation}}
\def\bea{\begin{eqnarray}}
\def\eea{\end{eqnarray}}
\def\bal{\begin{align}}
\def\eal{\end{align}}

\newcommand\sgn{\text{sgn}}

\title{Quantisation of super Teichm\"uller theory}

\author[a,b]{Nezhla Aghaei}
\author[a]{Michal Pawelkiewicz}
\author[a]{J\"org Teschner}

\affiliation[a]{DESY, Theory Group, Notkestrasse 85, Building 2a, 22607 Hamburg, Germany}
\affiliation[b]{Department of Mathematics, University of Hamburg, Bundesstrasse 55, 20146 Hamburg, Germany}

\emailAdd{nezhla.aghaei@desy.de}
\emailAdd{michal.pawelkiewicz@desy.de}
\emailAdd{joerg.teschner@desy.de}

\abstract{We construct a quantisation of the \TM spaces of super Riemann surfaces 
using coordinates associated to 
ideal triangulations of super Riemann surfaces. A new feature is the non-trivial dependence on the 
choice of a spin structure which can be encoded combinatorially in a certain refinement of the 
ideal triangulation. By constructing a projective unitary representation of the groupoid of 
changes of refined ideal triangulations we demonstrate that the dependence of the resulting quantum theory
on the choice of  a triangulation is inessential.
}

\begin{document} 

\maketitle
\flushbottom

\section{Introduction}
The quantum theories obtained by
quantisation of moduli spaces of flat 
connections on Riemann surfaces
are known to have deep connections with conformal field theory, 
quantum group theory, and the topology of three manifolds. A link
between these three subjects is provided by the Chern-Simons theories 
associated to compact groups $G$.

The picture becomes even richer if the holonomy of the flat connections
takes values in non-compact groups like $G=SL(2,\mathbb{R})$ or 
$G=SL(2,\mathbb{C})$. The relevant conformal field theories will then 
be non-rational, having continuous families of primary fields, see \cite{T14a} for a recent review of 
some of these relations, and \cite{D14,AK} for recent progress on Chern-Simons theory with 
a complex gauge group. 

More recently it was discovered that 
the quantum theories of flat connection capture profound 
non-perturbative information on ${\cal N}=2$-supersymmetric 
gauge theories, see  \cite{T14b} for a review. 
The expectation values of certain loop observables in 
four-dimensional ${\cal N}=2$-supersymmetric 
gauge theories
coincide with the expectation values of natural observables in the
quantum theory of moduli spaces of flat connections. 

In the case
which is currently best understood one is dealing with the 
connected component of the moduli space of flat $PSL(2,\mathbb{R})$-connections
which is isomorphic to the Teichm\"uller space of Riemann surfaces \cite{Go,Hi}.
The relevant observables then acquire an additional geometric
interpretation as (quantized) geodesic length function.
The corresponding conformal field theory is called Liouville theory. 
The study of Chern-Simons theories associated to non-compact groups
appears to be an extremely promising young field of research expected to
have various profound links with three-dimensional hyperbolic geometry.

The Teichm\"uller theory has an interesting and rich generalisation 
provided by the deformation theory of super Riemann surfaces.
Initially motivated by superstring perturbation theory, there has 
been a lot of research (reviewed in \cite{Witten12})
on the complex analytic theory of super-Teichm\"uller
spaces. There is a uniformisation theorem for super-Riemann surfaces,
describing super Riemann surfaces as quotients of the super
upper half plane by discrete subgroups of $OSp(1|2)$ \cite{Crane:1986uf} providing us with an
alternative picture on super \TM theory similar to the perspective on ordinary \TM theory offered
by hyperbolic geometry.
The theory of super Riemann surfaces should lead to interesting 
generalisations of two- and three-dimensional hyperbolic 
geometry, currently much less developed than the corresponding 
theories for ordinary Riemann surfaces. 

It should, in particular, be  interesting to develop the quantum 
theory of super Riemann surfaces. This may be expected to lead to a
new class of invariants of three-manifolds. 
It is furthermore known that there are  generalisations
of the relation between four-dimensional
${\cal N}=2$-supersymmetric 
gauge theories and conformal field theory  discovered by Alday, Gaiotto and Tachikawa \cite{AGT},
where
Super-Liouville theory appears instead of 
bosonic Liouville theory \cite{Belavin:2011pp}.
It seems likely that such generalisations are
related to the quantum 
theory of super Riemann surfaces in a way that 
is analogous to the relations between gauge theories, 
Liouville theory and the quantum Teichm\"uller theory \cite{T14b}.

In this paper we will develop the basic groundwork of the 
quantum theory of super Riemann surfaces. The approach
is similar to the the one used by Kashaev  in \cite{Kash1} for the case of 
ordinary \TM theory based on a suitable collection of coordinates
associated to the triangles forming an ideal 
triangulation of the surface.
An important new feature is the
dependence of the resulting theory on the choices of 
spin structures. Following the approach of Cimansoni and 
Reshetikhin \cite{cr1,cr2}, we will encode the choices of spin structures into
combinatorial data called Kasteleyn orientations, suitably 
adapted to the triangulations of our interest.

A basic issue to address in any approach based on triangulations
is to demonstrate the independence of the resulting quantum theory
on the choice of triangulation. This can be done by the constructing 
unitary operators relating the quantum theories associated to any 
two given triangulations. Being unitarily equivalent, one may
identify the quantum theories associated to two different 
triangulations as different representations of one and the same
quantum theory. The unitary operators representing changes of 
triangulations generate a projective representation of the super
Ptolemy groupoid describing the transitions between suitably refined  
triangulations equipped with Kasteleyn orientations. \\

The paper is organised as follows.
In the section \ref{chapter2} we review ordinary \TM theory and its quantisation. We discuss how to parametrise the \TM space using two sets of coordinates associated to triangulations which will have natural analogues in the case of
super \TM theory.
We then proceed to discuss the quantisation of this theory and the projective representation
of the Ptolemy groupoid relating the Hilbert spaces assigned to different triangulations. 

In the section \ref{chapter3} we discuss the super \TM theory. In order to encode the choices of spin structure
we will refine the triangulations into graphs called hexagonalisations. Such graphs with 
chosen Kasteleyn orientations can be used to define super analogues of the shear coordinates \cite{BB}. 
Changes of hexagonalisations define an analogue of the super Ptolemy groupoid which can be characterised in 
terms of generators and relations. 

The following section \ref{chapter5} describes the quantisation of the classical super \TM theory. We 
define operators representing analogues of the coordinates  used  in the work of Fock \cite{Fock} and 
Kashaev \cite{Kash1}, respectively, as well as the generators of the super Ptolemy groupoid 
describing changes of triangulations. The relations of the super Ptolemy groupoid follow
from  identities satisfied by suitable variants of Faddeev's quantum dilogarithm.

Section 5 finally offers an outlook.


\section{Ordinary \TM theory and its quantisation}\label{chapter2}

In order to prepare for the case of super \TM theory we will find it useful to briefly
review  relevant background on the \TM spaces of deformations of complex structures on 
Riemann surfaces. 
In the following we will consider two-dimensional surfaces $\Sigma_{g,n}$ 
with genus $g\geq 0$ and $n\geq 1$
punctures having $2g-2+n>0$. 
Useful starting points for 
the quantisation of the \TM spaces are the coordinates 
introduced by Penner \cite{Penner}, and their relatives
used in the works of Fock \cite{Fock}, Chekhov and Fock \cite{Chekhov:1999tn} and Kashaev \cite{Kash1}. 
Using these coordinates one may define an
an essentially canonical quantisation of the \TM spaces.

\subsection{Classical \TM theory}

\begin{figure}[h]
\centering
  \includegraphics[width=0.8\linewidth]{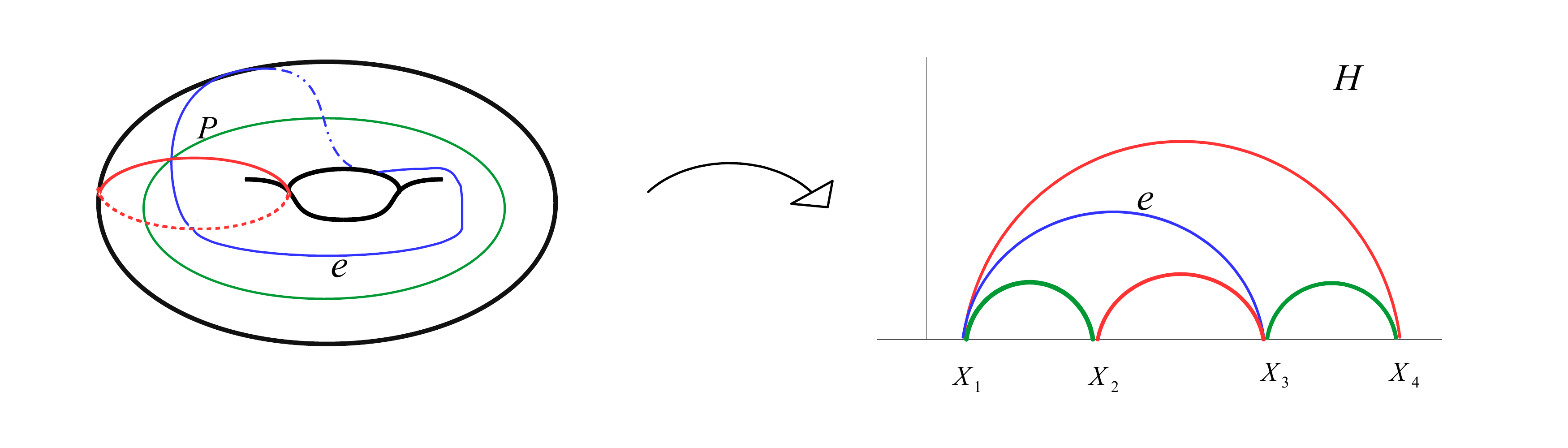}
  \caption{Realisation of a quadrilateral laying on a Riemann surface on upper half plane.}
   \label{uniformisation}
\end{figure}

 
 The uniformisation theorem states that Riemann surfaces $\Sigma_{g,n}$  can be represented as
 quotients of the upper half-plane $\mathbb{H}=\left\{ z\in \mathbb{C}:{\rm Im}(z)>0 \right\} $ equipped with the 
 Poincar\'e metric $ds^2=\frac{dyd\bar{y}}{({\rm Im}(y))^2}$ by 
 discrete subgroups $\Gamma$ of $PSL(2,\mathbb{R})$ called  Fuchsian groups\footnote{Discrete subgroups 
 of $PSL(2,\mathbb{R})$ having no elliptic elements.},  
\begin{equation}
 \Sigma_{g,n} \equiv \mathbb{H}\slash \Gamma .
\end{equation}
We may represent the points on $\Sigma_{g,n}$ as points in
a fundamental domain $D$ in the upper-half plane on which $\Gamma$ acts properly discontinuously.
The $n$ punctures of $\Sigma_{g,n}$ will be represented by a collection of points on the
boundary of $\mathbb{H}$ which can be identified with the projective real line $\mathbb{R}\mathbb{P}^1$.
Figure \ref{uniformisation} illustrates the uniformisation of a once-punctured torus $\Sigma_{1,1}$.

The \TM space $\mathcal{T}_{g,n}$ of Riemann surfaces $\Sigma_{g,n}$ 
can the be identified with the connected component in 
\begin{equation}
 \mathcal{T}_{g,n} = \{ \psi: \pi_1(\Sigma_{g,n}) \to PSL(2,\mathbb{R}) \} \slash PSL(2,\mathbb{R}).
\end{equation}
that contains all Fuchsian representations $\psi$.
The group $PSL(2,\mathbb{R})$ acts on representations $\psi$  by conjugation.


Useful sets of coordinates for the \TM spaces can be associated to ideal triangulations of $\Sigma_{g,n}$. 
Such a triangulation
can be defined by a maximal set of geodesic arcs intersecting only at the 
punctures of $\Sigma_{g,n}$ representing  their start- and endpoints. Such a 
collection of arcs decomposes the 
surface $\Sigma_{g,n}$ into a collection of triangles, as is also illustrated in figure \ref{uniformisation}.
An ideal triangulation $\tau$ of  Riemann surface $\Sigma_{g,n}$ is defined by 
 $3(2g-2+n)$ arcs, henceforth called edges, and has $2(2g-2+n)$ triangles. 

 


Useful coordinates may be assigned to the edges of an ideal triangulation by 
assigning to an edge $e$ separating two triangles as illustrated in figure \ref{uniformisation}
the cross-ratio 
\begin{equation}\label{conformalinvariant}
 e^{-z_e}=\frac{(x_1-x_2)(x_3-x_4)}{(x_1-x_4)(x_2-x_3)} ,
\end{equation}
formed out of the points $x_1,x_2,x_3,x_4$ representing the corners of the quadrilateral 
decomposed into
two triangles by the edge $e$.
The resulting set of $6g-6+3n$ coordinate functions may be used to get a system of coordinates for 
\TM space by taking into account the relations
$\sum_{e\in E(P_i)} z_{e} = 0$, where $E(P)$ is the set of edges ending in puncture $P$.


The Poisson structure on $\mathcal{T}_{g,n} $ defined by the Weil-Petersson symplectic form
takes a particularly simple form in the coordinates $z_e$. It may be represented as
\begin{equation} \label{fockpoissonbracket}
\{ z_e,z_f\}_{WP}^{}= n_{ef}\,,
\end{equation}
where $n_{ef}$ is the number of times $e$ and $f$ meet in a common end-point $P$, counted positively
if $f$ is the first edge reached from $e$ upon going around $P$ in clockwise direction, counted negatively otherwise.
%
%
%

\begin{figure}[h]
\centering
\includegraphics[width=0.8\textwidth]{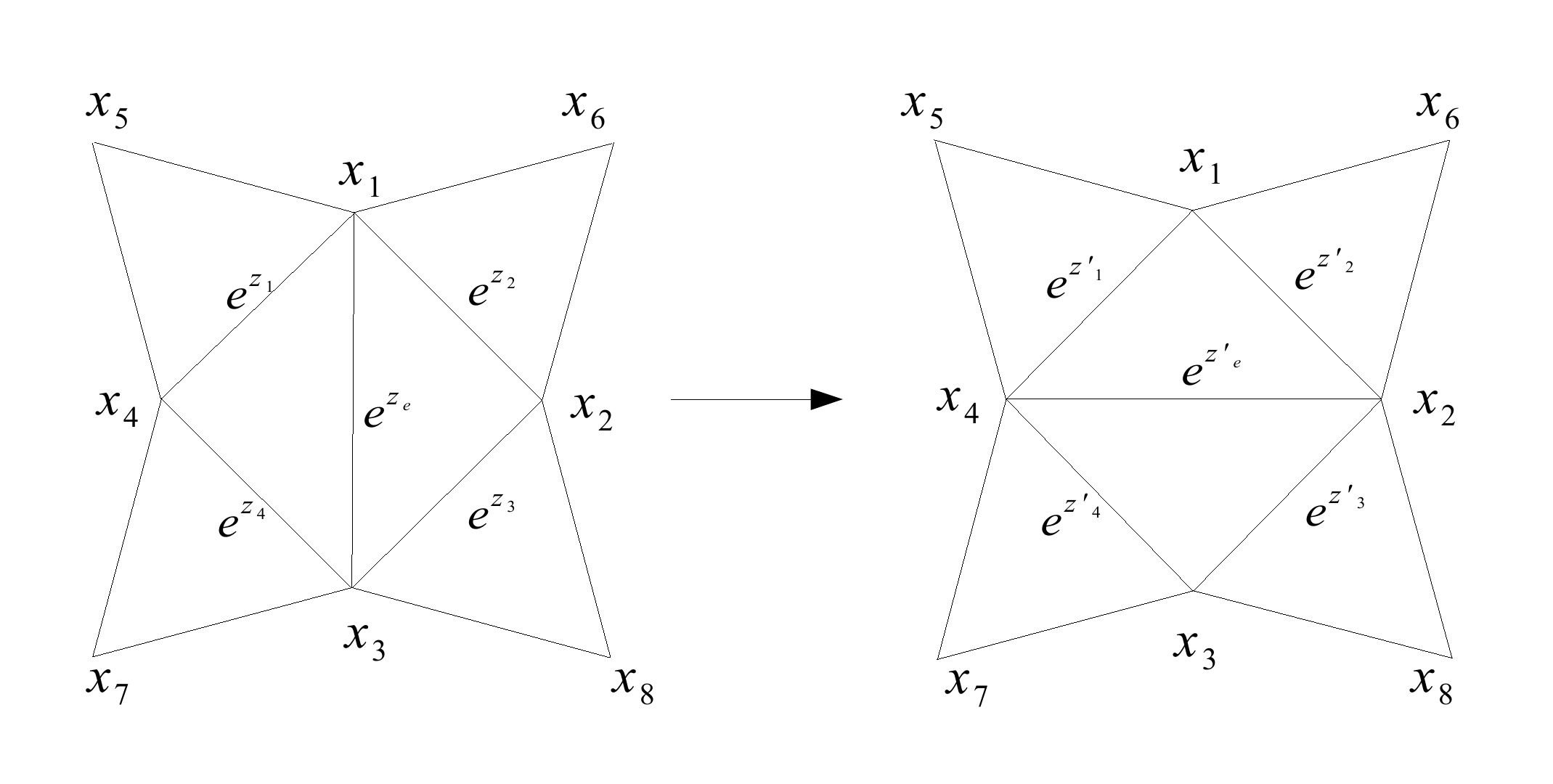}
\caption{Flip of an ideal triangulation.}
\label{map}
\end{figure}

The definition of the shear coordinates $z_e$ was based on the choice of an ideal triangulation. 
Changing the ideal triangulation defines new coordinates $z_e'$ that can be expressed in terms
of the coordinates $z_e$. General changes of triangulation can be represented as compositions of 
the elementary operation called flip changing the diagonal in one quadrilateral only, as illustrated in 
figure \ref{map}.
This change of triangulation induces the following change of coordinates 
\begin{equation}\label{sheartrsf}
\begin{aligned}
 &e^{z'_1} = e^{{z_1}}(1 + e^{z_e}),\\
 &e^{z'_4} = e^{{z_4}}(1 + e^{-z_e})^{-1},
 \end{aligned}\qquad
 e^{z'_e} = e^{-z_e},\qquad
 \begin{aligned}
  &e^{z'_2} = e^{{z_2}}(1 + e^{-z_e})^{-1},\\
 &e^{z'_3} = e^{{z_3}}(1 + e^{z_e}).
\end{aligned}
\end{equation}
leaving all other coordinates unchanged.

\subsection{Kashaev coordinates}\label{bos-kash}

As a particularly useful starting point for quantisation it has turned out to be useful to describe the Teichm\"uller spaces 
by means of a set of coordinates associated to the triangles rather than the edges of an ideal triangulation 
\cite{Kash1}.
We shall
label the triangles $\Delta_v$ by $v=1,\dots,4g-4+2n$ and choose a distinguished corner in every one of them.
One may then assign to each triangle a pair of
variables $(p_v,q_v)$ allowing us to recover the variables $z_e$ as
\begin{equation} \label{ze}
z_e=\tilde{z}_{e,v}+\tilde{z}_{e,w},\qquad 
\tilde{z}_{e,v}= \begin{cases}     
    p_v \quad &\text{if} \quad e=e_1^v,\\
    -q_v \quad &\text{if} \quad e=e_2^v,\\
    q_v-p_v \quad &\text{if} \quad e=e_3^v,
\end{cases}
\end{equation}
where $e_i^v$ are the edges surrounding triangle $\Delta_v$ counted by $i=1,2,3$ in counter-clockwise order
such that $e_i^3$ is opposite to the distinguished corner. 

The space $\mathbb{R}^{4(2g-2+n)}$ will be equipped with a Poisson structure defined by
\begin{equation}
 \begin{aligned}
  \{p_v,p_w\} &= 0, \\
  \{q_v,q_w\} &= 0, 
 \end{aligned}\qquad   \{p_v,q_w\} = \delta_{v,w}.
\label{kashaevpoissonbracket} 
\end{equation}
It can be shown that the Poisson structure on Kashaev coordinates given by \eqref{kashaevpoissonbracket} induces the Poisson structure on shear coordinates \eqref{fockpoissonbracket} via \eqref{ze}.

One may then describe the Teichm\"uller space using the Hamiltonian reduction of $\mathbb{R}^{4(2g-2+n)}$
with Poisson bracket (\ref{kashaevpoissonbracket}) with respect to a suitable set of constraints $h_\gamma$
labelled by $\gamma\in H_1(\Sigma_{g,n},\mathbb{Z})$, and represented 
as linear functions in the $(p_v,q_v)$ \cite{Kash1}. The functions $z_e$ defined via (\ref{ze}) satisfy 
$\{h_\gamma,z_e\}=0$ for all edges $e$ and all $\gamma\in H_1(\Sigma_{g,n},\mathbb{Z})$ and may therefore be used 
to get coordinates for the subspace defined by the constraints.

\begin{figure}[h] 
\centering
\includegraphics[width=0.5\textwidth]{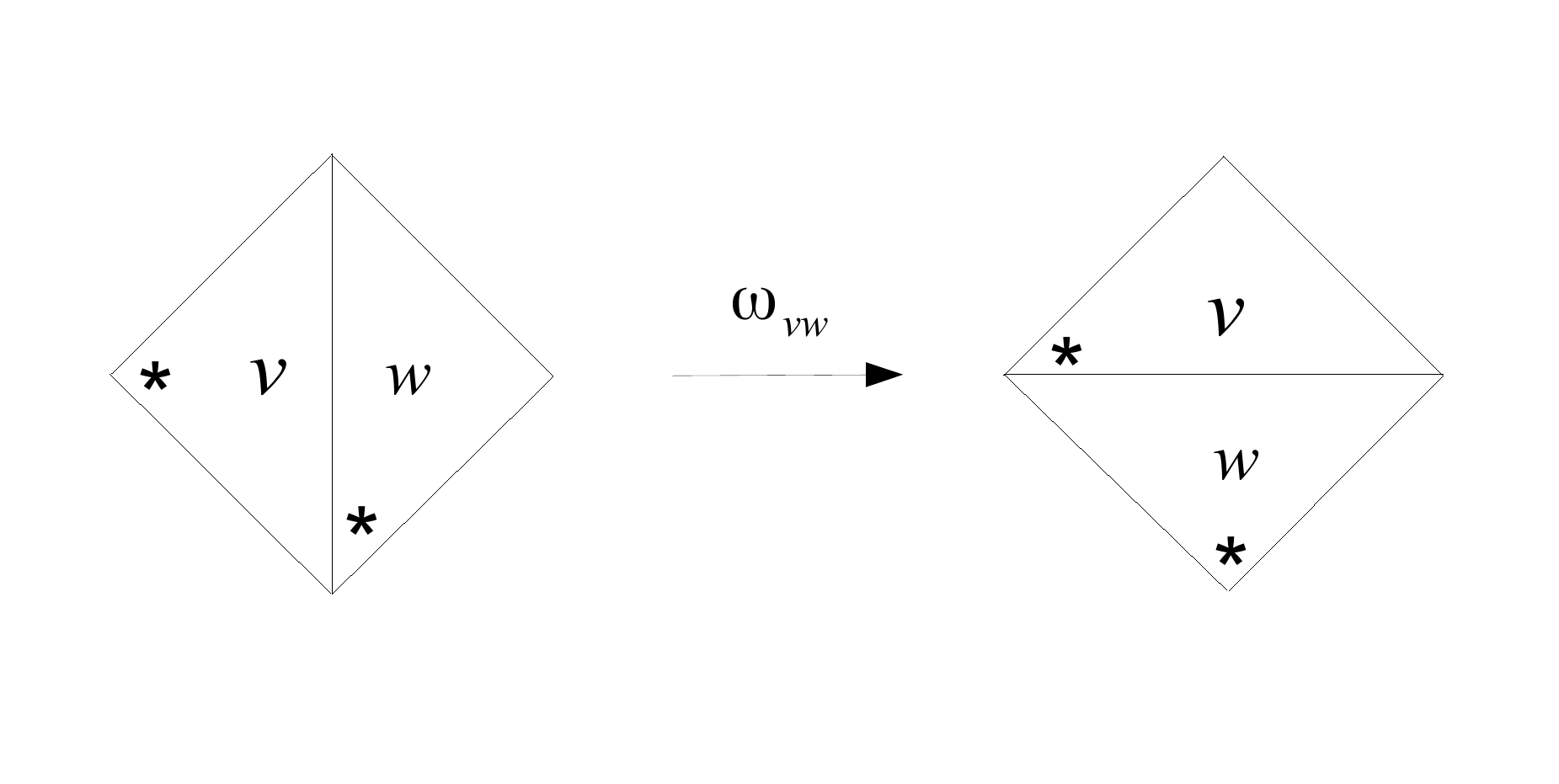}
\caption{The flip transformation $\omega_{vw}$.}
\label{classicalmap1}
\end{figure}

\begin{figure}[h] 
\centering
\includegraphics[width=0.5\textwidth]{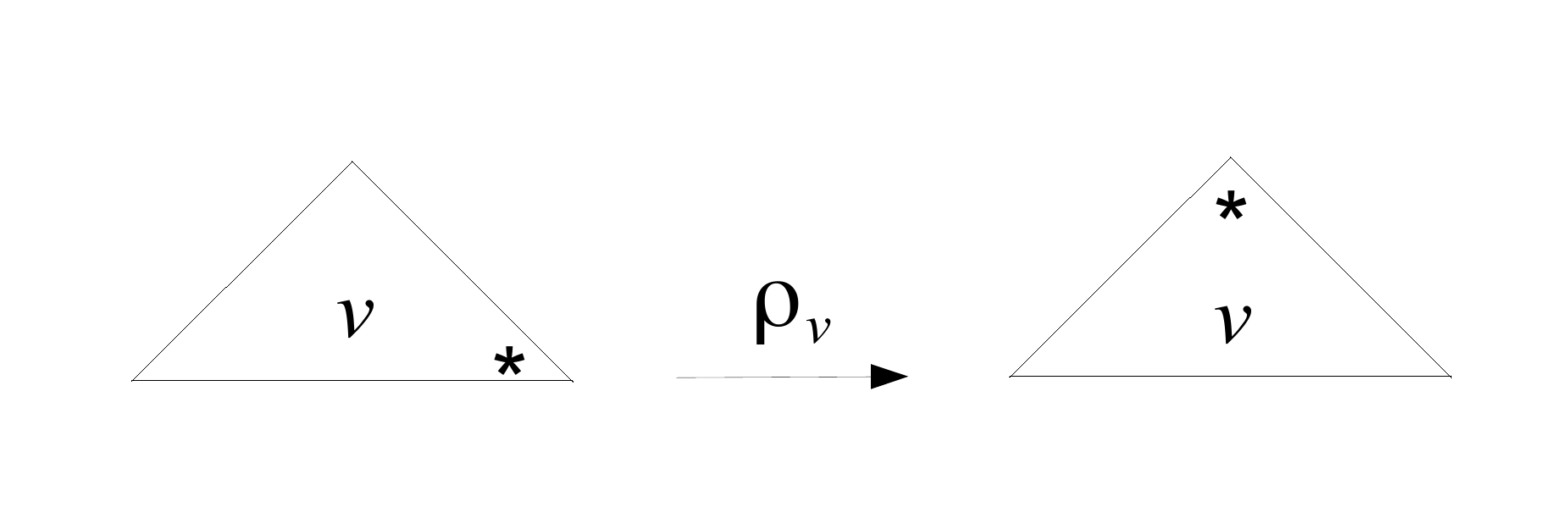}
\caption{The transformation $\rho_v$.}
\label{classicalmap2}
\end{figure}

One may define changes of Kashaev coordinates associated to any changes of ideal triangulations
preserving the Poisson structure, and inducing the changes of shear coordinates (\ref{sheartrsf})
via (\ref{ze}). Having equipped the ideal triangulations with an additional decoration represented by the 
numbering of the triangles $\Delta_v$ and the choice of a distinguished corner in each triangle forces
us to consider an enlarged set of elementary transformations relating arbitrary decorated 
ideal triangulations. Elementary transformations 
are the flips $\omega_{vw}$, the rotations $\rho_v$ and the permutations $(vw)$.
Flips $\omega_{vw}$ and rotations $\rho_v$ are illustrated in 
figures \ref{classicalmap1} and \ref{classicalmap2}, respectively, while the 
permutation $(uv)$ simply exchanges the labels of the triangles $u$ and $v$.

The change of coordinates associated to the transformation
$\rho_v$ is given as
\begin{equation}
 \rho_v^{-1}:(q_v,p_v)\rightarrow(p_v-q_v,-q_v),
\end{equation}
while under a flip $\omega_{vw}$ the transformation of Kashaev coordinates is realised by  
\begin{equation}\label{kashaevflipclassical}
\omega_{vw}^{-1}: \begin{cases}     
    (U_v,V_v)\rightarrow(U_vU_w,U_vV_w+V_v), \\
   (U_w,V_w)\rightarrow(U_wV_v(U_vV_w+V_v)^{-1},V_w(U_vV_w+V_v)^{-1}),
\end{cases}
\end{equation}
where we denote $U_v\equiv e^{q_v}$ and $V_v=e^{p_v}$.



\begin{figure}[h]\label{classicalpentagonexample}
\centering
\includegraphics[width=0.7\textwidth]{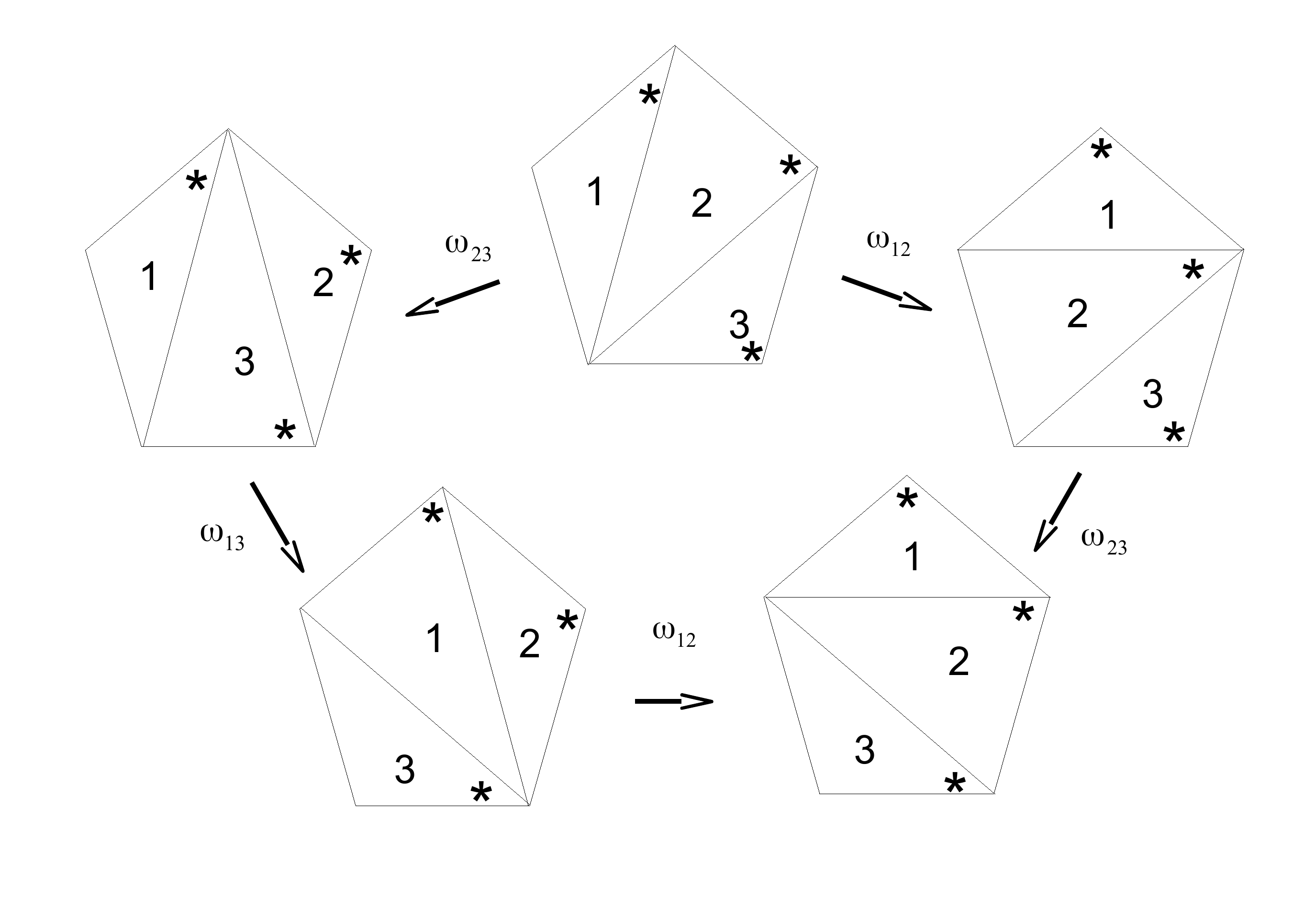}
\caption{The pentagon equation.}
\label{classicalpentagon}
\end{figure}

The 
transformations between decorated ideal triangulations generate a groupoid that can be described 
in terms of generators and relations.
As we mentioned above, any two decorated triangulations of the same Riemann surface can be related by a finite sequence of permutations $(vw)$, flips $\omega_{vw}$ and rotations $\rho_v$. 
Any  sequence of elementary transformations returning to its initial point
defines a relation. A basic set of relations implying all others is known to be the following
\begin{subequations}
\begin{align}
 & \rho_v \circ \rho_v \circ \rho_v = id_v, \label{rel1}\\
 & (\rho_v^{-1}\rho_w) \circ \omega_{vw} = \omega_{wv} \circ (\rho_v^{-1} \rho_w), \\
 & \omega_{wv} \circ \rho_v \circ \omega_{vw} = (vw) \circ (\rho_v\rho_w) ,\label{rel3} \\
 & \omega_{v_1 v_2} \circ \omega_{v_3 v_4} = \omega_{v_3 v_4} \circ \omega_{v_1 v_2}, \quad v_i\neq v_j, i\neq j,\\
 & \omega_{v w} \circ \omega_{u w} \circ \omega_{u v} = \omega_{u v} \circ \omega_{v w} ,  \label{pentagon}
\end{align}
\label{classicalptolemyrelations}
\end{subequations}
The pentagon relation (\ref{pentagon}) illustrated in figure \ref{classicalpentagon} is of particular importance,
while the relations  \eqref{rel1}-\eqref{rel3} describe changes of the decorations.

\subsection{Quantum \TM theory}
 
  Quantisation of the \TM theory of punctured Riemann surfaces was developed by Kashaev in \cite{Kash1} and independently by Fock and Chekhov in  \cite{Fock,Chekhov:1999tn}.
 We will associate a Hilbert space $\mathcal{H}_v = L^2(\mathbb{R})$ with each triangle of a decorated triangulation.
 The Hilbert space associated to the entire triangulation is the tensor product 
 \begin{equation}
  \mathcal{H} = \bigotimes_{v=1}^{4g-4+2n} \mathcal{H}_v.
 \end{equation}
 In the quantum theory one replaces the coordinate functions $(p_v,q_v)$ 
 by self-adjoint operators $(\mathsf{p}_v,\mathsf{q}_v)$, ${v=1,\dots,4g-4+2n}$,
having the following commutation relations
 \begin{equation}
   \left[\mathsf{p}_v,\mathsf{q}_w\right]=\frac{1}{2\pi i}\delta_{vw}, \qquad
  \begin{aligned}
    \left[\mathsf{q}_v,\mathsf{q}_w\right]&= 0, \\
   \left[\mathsf{p}_v,\mathsf{p}_w\right]&=0 .
  \end{aligned}
 \end{equation}
Formula (\ref{ze}) has an obvious counterpart in the quantum theory, defining 
 self-adjoint operators  $\mathsf{z}_e$ having the relations
 \begin{equation}
 \left[  \mathsf{z}_e,\mathsf{z}_{e'} \right] = \frac{1}{2\pi i} \left\lbrace z_e,z_{e'}\right\rbrace _{WP} .
 \end{equation} 
 A quantum version of the Hamiltonian reduction procedure can be defined 
describing Hilbert space and algebra of observables  of the quantum theory of 
Teichm\"uller spaces in terms of the 
quantum theory defined above, 
see \cite{Kash1,Kash2,T05} for more details.


We finally need to describe the
quantum realisation of maps changing the triangulation $\tau$ of a Riemann surface $\Sigma$. The move $\rho_v$
rotating the distinguished vertex of a triangle $v$ is realised by an operator $\mathsf{A}_v: \mathcal{H}_v\to\mathcal{H}_v$ 
\begin{equation}
 \mathsf{A}_v = e^{i\pi/3} e^{-i3\pi \mathsf{q}_v^2}e^{-i\pi(\mathsf{p}_v+\mathsf{q}_v)^2}.
\end{equation}
The flips get represented by unitary operators 
$\mathsf{T}_{vw}:\mathcal{H}_v\otimes \mathcal{H}_w \to \mathcal{H}_v\otimes \mathcal{H}_w$ defined as
\begin{equation}
 \mathsf{T}_{vw} = e_\ub(\mathsf{q}_v+\mathsf{p}_w-\mathsf{q}_w)e^{-2\pi i \mathsf{p}_v \mathsf{q}_w},
\end{equation}
where $b$ is a parameter such that 
Planck's constant $\hbar = 2\pi b^2$, and $e_\ub$ is a quantum dilogarithm function defined 
in Appendix \ref{appendix1}. The quantised version of the transformation of the shear
coordinates takes the form
\begin{equation}\label{kashaevflipquantum}
 \begin{split}
  & \mathsf{T}^{-1}_{vw} e^{2\pi b \mathsf{z}_1'} \mathsf{T}^{}_{vw} = 
  e^{\pi b \mathsf{z}_1} (1 + e^{2\pi \mathsf{z}_e}) e^{\pi b \mathsf{z}_1}, \\
  & \mathsf{T}^{-1}_{vw} e^{2\pi b \mathsf{z}_2'} \mathsf{T}^{}_{vw} = e^{\pi b \mathsf{z}_2} (1 + e^{-2\pi \mathsf{z}_e})^{-1} e^{\pi b \mathsf{z}_2}, \\
  & \mathsf{T}^{-1}_{vw} e^{2\pi b \mathsf{z}_3'} \mathsf{T}^{}_{vw} = e^{\pi b \mathsf{z}_3} (1 + e^{2\pi \mathsf{z}_e}) e^{\pi b \mathsf{z}_3}, \\
  & \mathsf{T}^{-1}_{vw} e^{2\pi b \mathsf{z}_4'} \mathsf{T}^{}_{vw} = e^{\pi b \mathsf{z}_4} (1 + e^{-2\pi \mathsf{z}_e})^{-1} e^{\pi b \mathsf{z}_4}, \\
  & \mathsf{T}^{-1}_{vw} e^{2\pi b \mathsf{z}_e'} \mathsf{T}^{}_{vw} = e^{-2\pi b \mathsf{z}_e} ,
 \end{split}
\end{equation}
assuming that $\mathsf{T}^{}_{vw}$ represents the flip depicted in figure \ref{map} with decoration introduced in 
Figure \ref{classicalmap1}.
The equations \eqref{kashaevflipquantum} provide the quantisation of 
\eqref{kashaevflipclassical}, and we can recover the classical transformation by taking the limit $q=e^{i\pi b^2}\to 1$. 


The operators $\mathsf{T}_{uv}$ and $\mathsf{A}_v$ generate a projective representation of the 
Ptolemy groupoid characterised by the set of relations
\begin{align}
 &\mathsf{A}_1^3={\rm id}_1,\\
 &\mathsf{T}_{23} \mathsf{T}_{13} \mathsf{T}_{12} =\mathsf{T}_{12} \mathsf{T}_{23},\\
 &\mathsf{A}_2 \mathsf{T}_{12} \mathsf{A}_{1}= \mathsf{A}_1 \mathsf{T}_{21}\mathsf{A}_2,\\
 &\mathsf{T}_{21} \mathsf{A}_1 \mathsf{T}_{12}=\zeta \mathsf{A}_1\mathsf{A}_2\mathsf{P}_{(12)},
\end{align}\label{permutationequation}
where $\zeta=e^{\pi i {c_b}^2/3 }$ and $c_b=\dfrac{i}{2}(b+b^{-1})$. The permutation $\mathsf{P}_{(12)}: \mathcal{H}_1\otimes\mathcal{H}_2 \to \mathcal{H}_1\otimes\mathcal{H}_2$ is defined as the 
operator acting as
$\mathsf{P}_{(12)} (v_1\otimes v_2) = v_2\otimes v_1$ for all $v_i\in\mathcal{H}_i$.
 
The quantised flip transformation has an interesting relation with quantum group theory. Kashaev \cite{Kash3} has shown that one can identify the flip operator $\mathsf{T}$ with the canonical element of the Heisenberg double of the quantum plane, the Borel half of $U_q(sl(2))$, evaluated on particular infinite-dimensional representations. Moreover, the rotation 
operator $\mathsf{A}_v$ is an algebra automorphism of this Heisenberg double.

 
\section{Classical super Teichm\"uller spaces}\label{chapter3}

The aim of this chapter is to present the basics of super \TM theory, the \TM theory of super Riemann surfaces. Of particular importance will be 
the  coordinates for the super \TM spaces introduced in \cite{BB}.
These coordinates are closely related to the analogue of Penner's coordinates recently introduced in 
\cite{Penner:2015xla}. 
 
 In the following section 
we will, following \cite{BB} closely, 
first review the basic notions of super Riemann surfaces and super \TM spaces. 
We will then consider the definition of two sets of coordinates on this space. 
In order to define such coordinates we will need to refine the triangulations
used to define coordinates for the ordinary \TM spaces into certain graphs called 
hexagonalisations. 
Assigning the so-called Kasteleyn orientations to the edges of a hexagonalisation
allows one to 
parametrise the choices of spin structures on super Riemann surfaces. In addition to 
even coordinates associated to edges of the underlying triangulation one may define 
additional odd coordinates associated to the triangles. The additional orientation data assigned to a 
hexagonalisation are used to provide an unambiguous definition of the signs of the odd coordinates.

We will furthermore discuss
the transformations of coordinates induced by changes of hexagonalisations. 
The result of the elementary operation of changing the diagonal in a quadrangle called flip will now depend
on the choice of Kasteleyn orientation. We will furthermore need to consider 
an additional operation relating different hexagonalisations called push-out. 
This operation relates different Kasteleyn orientations describing the same spin structure.
The relations that have to be satisfied by these transformations define 
a generalisation of Ptolemy groupoid that will be called super Ptolemy groupoid.

\subsection{The super upper half plane and its symmetries}\label{invariants}

We will begin by introducing the basic group-theoretic 
and geometric background for the definition of the super \TM spaces
and for constructing convenient coordinates on these spaces. 

The coordinates on the two-dimensional super-plane $\mathbb{R}^{2|1}$
can be assembled in column or row-vectors $(x_1,x_2|\theta)$ with $x_i\in\mathbb{R}$, $i=1,2$,
and $\theta$ being an element of a Grassmann algebra satisfying $\theta^2=0$. 
The elements of 
the subgroup $OSp(1|2)$ of the group of linear transformations of $\mathbb{R}^{2|1}$ may be represented
by $(2|1)\times(2|1)$ matrices of the form
\begin{equation}
g= \left( \begin{array}{ccc}
         a & b & \gamma \\
         c & d & \delta \\
         \alpha & \beta & e
        \end{array}
 \right) ,
\end{equation}
when the matrix elements are elements of a Grassmann algebra satisfying the relations
\begin{align}\label{ospconditions1}
 & ad-bc-\alpha\beta=1,\\
 & e^2+2\gamma\delta=1,\\
 & \alpha e=a\delta-c\gamma, \\
 & \beta e=b\delta-d\gamma.\label{ospconditions2}
\end{align}
A natural map from  $OSp(1|2)$ to $SL(2,\mathbb{R})$ may be defined by 
mapping the odd generators to zero. The image of 
$g\in OSp(1|2)$  under this map will be denoted as $g^{\sharp}\in SL(2,\mathbb{R})$.

The super upper half-plane is defined as
$\mathbb{H}^{1|1}=\left\{ (z,\theta)\in \mathbb{C}^{1|1}:{\rm Im}(z)>0 \right\} $. 
 $OSp(1|2)$ acts on the super upper half plane $\mathbb{H}^{1|1}$ by generalised M\"obius transformations of  
 the form
\begin{align}
&z \longrightarrow z'=\frac{az+b+\gamma \theta}{cz+d+\delta\theta}, \\
&\theta \longrightarrow \theta'=\frac{\alpha z+\beta+e \theta}{cz+d+\delta\theta}\,.
\end{align}
The one-point compactification of the boundary of $\mathbb{H}^{1|1}$ is the super projective real line
denoted 
by $\mathbb{P}^{1|1}$. Elements of $\mathbb{P}^{1|1}$ may be represented as 
column or row vectors $(x_1,x_2|\theta)$ with $x_i\in\mathbb{R}$, $i=1,2$ 
modulo overall multiplication by non-vanishing real numbers. 
Considering 
vectors $(x_1,x_2|\theta)$ with $x_i\in\mathbb{R}$, $i=1,2$ 
modulo overall multiplication by non-vanishing {\it positive}  numbers defines a
double cover $\mathbb{S}^{1|1}$ of $\mathbb{P}^{1|1}$.

There are two types of invariants generalising the cross-ratio present in the ordinary case.
To a collection of four points with coordinates $P_i=({x_i | \theta_i})$, $i=1,\dots,4$  
one may assign a super-conformal cross-ratio
\begin{equation}\label{conformalinvariantt}
 e^{-z}=\frac{X_{12}X_{34}}{X_{14}X_{23}} ,
\end{equation}
where $X_{ij} = x_i-x_j-\theta_i\theta_j$. To a collection of three
points $P_i=({x_i | \theta_i})$, $i=1,\dots,3$ one may furthermore
be tempted to assign
an odd (pseudo-) invariant via
\begin{equation}\label{oddinv}
 \xi =  \pm \frac{x_{23}\theta_1+x_{31}\theta_2+x_{12}\theta_3 - \frac{1}{2}\theta_1\theta_2\theta_3}{(X_{12}X_{23}X_{31})^\frac{1}{2}},
\end{equation}
where $ x_{ij}=x_i-x_j$. Due to the appearance of a square-root one can use the expression in (\ref{oddinv})
to define $\xi$ up to a sign.

In order to arrive at an unambiguous definition
one needs to fix a prescription for the definition of the sign of $\xi$. A convenient way to parametrise the
choices involved in the definition of the odd invariant uses the so-called Kasteleyn orientations of the
triangles in $\mathbb{H}^{1|1}$ with corners at $P_i$, $i=1,\dots,3$. A Kasteleyn orientation of a polygon
embedded in an oriented surface is an orientation for the sides of the polygon
such that the number of sides oriented against  the induced orientation on the boundary of the 
polygon is odd.

A Kasteleyn orientation of triangles with three corners at $P_i\in\mathbb{P}^{1|1}$, $i=1,\dots,3$ 
may then by used to define lifts of the points $P_i\in\mathbb{P}^{1|1}$ to 
points $\hat{P}_i$ of its double cover $\mathbb{S}^{1|1}$ for $i=1,2,3$ as follows. We may choose an arbitrary lift of 
$P_1$, represented by a vector $(x_1,y_1|\theta_1)\in \mathbb{R}^{2|1}$. If the edge connecting $P_i$ to $P_1$
is oriented from $P_1$ to $P_i$, $i=2,3$, we will choose lifts of $P_i$ represented by vectors
$(x_i,y_i|\theta_i)\in \mathbb{R}^{2|1}$ 
such that ${\rm \sgn}\big({\rm \det}\big(\begin{smallmatrix} x_1 & x_i \\ y_1 & y_i\end{smallmatrix}\big)\big)=-1$,
while in the other case $P_i$ will be represented by vectors 
$(x_i,y_i|\theta_i)\in \mathbb{R}^{2|1}$ satisfying
${\rm \sgn}\big({\rm \det}\big(\begin{smallmatrix} x_1 & x_i \\ y_1 & y_i\end{smallmatrix}\big)\big)=1$.
By means of $OSp(1|2)$-transformations one may then map $\hat{P}_i$, $i=1,2,3$ to 
a triple of points $Q_i$ of the form $Q_1\simeq(1,0|0)$, $Q_3\simeq(0,-1|0)$, and $Q_2\simeq\pm (1,-1|\xi)$.
This allows us to finally define the odd invariant associated to a triangle with corners $P_i$, $i=1,2,3$, 
and chosen Kasteleyn orientation of its sides
to be equal to $\xi$ if $Q_2\simeq (1,-1|\xi)$,
and equal to $-\xi$ if $Q_2\simeq -(1,-1|\xi)$.

\subsection{Super Riemann surfaces and super Teichm\"uller space}

For our goals it will be most convenient to simply define super Riemann surfaces as 
quotients of the super upper half plane by suitable discrete subgroups of $\Gamma$ of $OSp(1|2)$.
This approach is related to the complex-analytic point of view reviewed in \cite{Witten12} by
an analogue of the uniformisation theorem proven in \cite{Crane:1986uf}.

A discrete subgroup of $\Gamma$ of $OSp(1|2)$ 
such that $\Gamma^\sharp$ is a Fuchsian group is called a super Fuchsian group. 
Super Riemann surfaces will be defined as 
quotients of the super upper half-plane $\mathbb{H}^{1|1}$ by a super Fuchsian group $\Gamma$,
\begin{equation}
 \Sigma_{g,n} \equiv \mathbb{H}^{1|1}\slash \Gamma .
\end{equation}
The points of a super Riemann surface may be represented by the points of a fundamental 
domain 
$D$ on the super upper-half plane on which $\Gamma$ acts properly discontinuous. 
Super Riemann surfaces with $n$ punctures have fundamental 
domains $D$ touching the boundary $\mathbb{P}^{1|1}$ of $\mathbb{H}^{1|1}$ in $d$ 
distinct points $P_i$, $i=1,\dots,d$.\footnote{When pairs of 
points get identified by the action of the group $\Gamma$ we will have $d\neq n$.}

We can finally define the super \TM space $\mathcal{ST}_{g,n}$ of super Riemann surfaces $\Sigma_{g,n}$ of genus $g$ with $n$ punctures. It can be represented  as the quotient
\begin{equation}
 \mathcal{ST}_{g,n} = \big\{ \rho: \pi_1(\Sigma_{g,n}) \to OSp(1|2) \big\}\, \slash \,OSp(1|2),
\end{equation}
where $\rho$ is a discrete 
representation of fundamental group $\pi_1(\Sigma_{g,n})$ into $OSp(1|2)$ whose image is super Fuchsian.
 
 There is always an ordinary Riemann surface $\Sigma_{g,n}^\sharp$ associated to each super Riemann
surface, defined as quotient of the upper half plane $\mathbb{H}$ by $\Gamma^\sharp$. 
Notions like ideal triangulations will therefore have obvious counterparts in the theory of 
super Riemann surfaces.

\subsection{Hexagonalisation and Kasteleyn orientations}

Similarly to the ordinary \TM spaces, the  parametrisation of super \TM spaces
introduced in \cite{BB} relies on  ideal 
triangulations of super Riemann surfaces. It will be based on the even and odd invariants of the group 
$OSp(1|2)$ that we defined in Section \ref{invariants}. However, as noted there, one need to introduce 
additional data to define the odd invariants unambiguously. The extra data must allow us to define the
lifts of the punctures $P_i\in\mathbb{P}^{1|1}$ to 
points $\hat{P}_i$ on its double cover $\mathbb{S}^{1|1}$. Note that the 
even part of $\mathbb{P}^{1|1}$ is the real projective line $\mathbb{R}\mathbb{P}^1$ with group of automorphisms
$PSL(2,\mathbb{R})$, while the even part of $\mathbb{S}^{1|1}$ is a double cover of $\mathbb{R}\mathbb{P}^1$
with group of automorphisms $SL(2,\mathbb{R})$. Lifting the vertices of a triangulation of $\mathbb{H}^{1|1}$ 
to $\mathbb{S}^{1|1}$ should therefore be accompanied with a lift of the Fuchsian group $\Gamma^\sharp\subset PSL(2,\mathbb{R})$
to a subgroup of $SL(2,R)$.  It is known that the definition of such a lift depends on the choice of 
a spin structure on $\Sigma$ \cite{natanzon}. We therefore need to introduce a suitable refinement of an ideal triangulation
which will allow us to encode the extra data defining a spin structure.



\begin{figure}[h]
\centering
\includegraphics[width=0.9\textwidth]{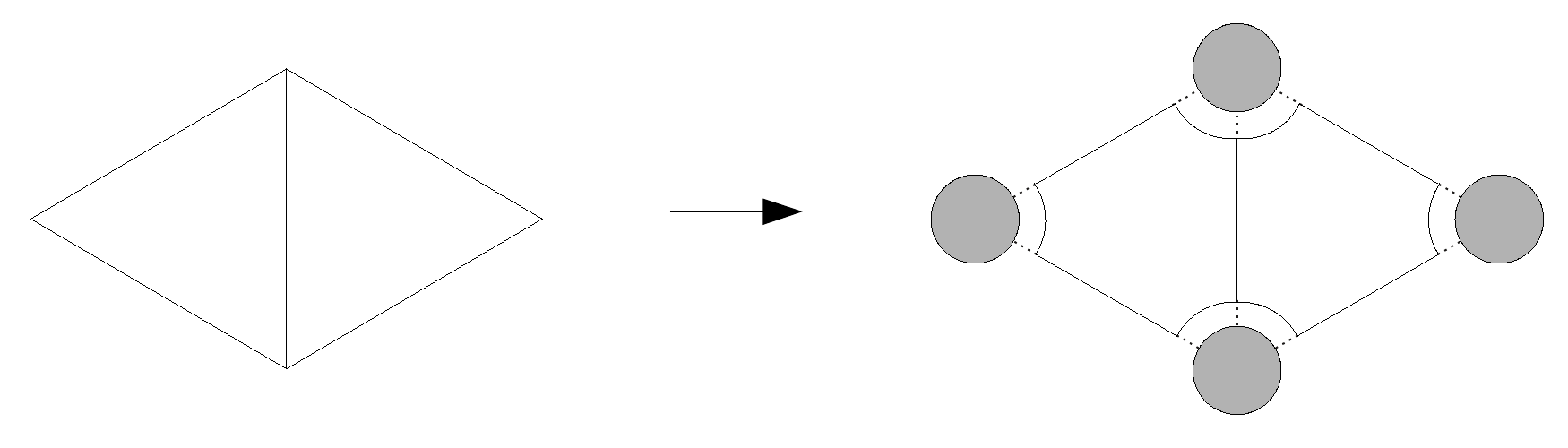}
\caption{Hexagonalisation.}
\label{dischexagon}
\end{figure}

The parametrisation of spin structure on Riemann surfaces used in \cite{BB} is based on results of 
Cimasoni,  Reshetikhin \cite{cr1,cr2} using Kasteleyn orientations. To begin with, let us first
introduce the notion of  a hexagonalisation. The starting point will be an ideal triangulation of a 
surface $\Sigma$.
Around each puncture let us cut out a small disc,
giving a surface $\Sigma_{\rm b}$ with $n$ holes. 
The parts of any two edges bounding a triangle in $\Sigma$ which are contained in $\Sigma_{\rm b}$
will then be connected by an arc in the interior of
$\Sigma_b$. The resulting hexagon has a boundary consisting of "long" edges coming from the 
edges of the original triangulation, and "short" edges represented by the arcs connecting the long edges.
The procedure is illustrated  in  figure \ref{dischexagon}.

Let us finally  introduce another set of edges called dimers connecting the
vertices of the hexagons with the boundary of $\Sigma_{\rm b}$. The dimers 
are represented by dashed lines in Figure \ref{dischexagon}.
The resulting graph will be called a hexagonalisation of the given ideal  triangulation.

The next step is to introduce a Kasteleyn orientation on the hexagonalisation defined above. It is 
given by an orientation of the boundary edges of the hexagons such that for every face of the 
resulting graph the number of edges oriented against the orientation of the surface is odd. 
It then follows from Theorem 1 in \cite{cr2} that the choice of the spin structure can be 
encoded in the choice of a Kasteleyn orientation on a hexagonalisation.\footnote{
The hexagonalisations constructed above are special cases of what is called surface graph with 
boundary in \cite{cr1,cr2}.  The formulation 
of Theorem 1 in \cite{cr2} makes use of the notion of a dimer configuration on a surface graph with 
boundary. In our case the dimer configuration is given by the set of edges connecting the corners of the
hexagons with the boundary shown as dashed lines in Figure \ref{dischexagon}.}

\begin{figure}[h]
\centering
\includegraphics[width=0.7\textwidth]{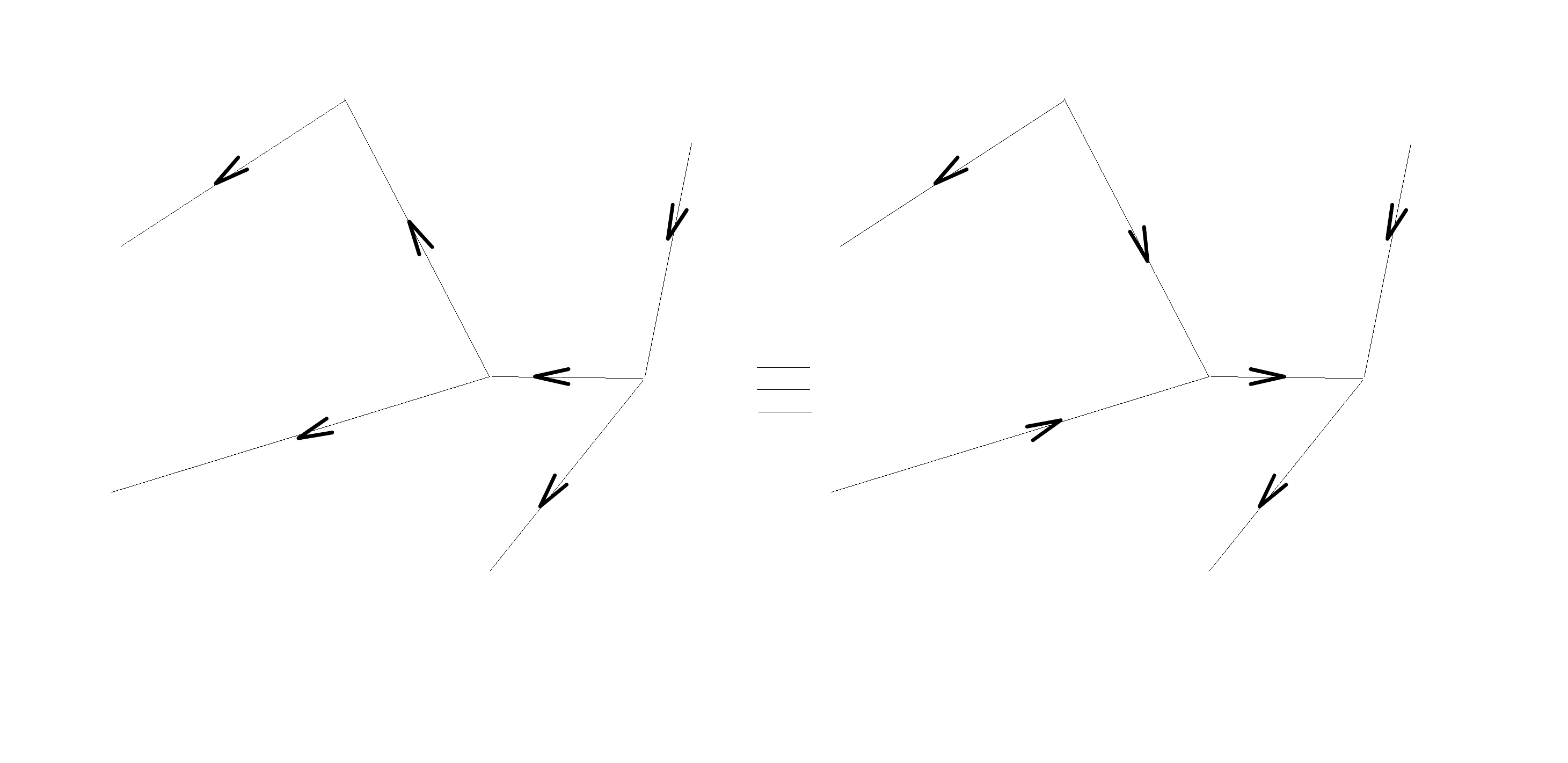}
\caption{Equivalence between the Kasteleyn orientations.}
\label{equalkastlegs}
\end{figure}

Different Kasteleyn orientations may describe the same spin structure. Two Kasteleyn orientations
are equivalent in this sense if they are
 related by the reversal of orientations of all the edges meeting at 
 the same vertex, as illustrated in Figure \ref{equalkastlegs}. 
The equivalence classes of Kasteleyn orientations related by this operation are in one-to-one correspondence
to the spin structures on $\Sigma$.

In order 
to represent a hexagonalisation with Kasteleyn orientation graphically we will find it convenient 
to contract all short edges to points, and marking the corners of the resulting triangle 
coming from short edges with orientation 
against the orientation of the underlying surface by dots. An illustration of this procedure is given 
in Figures \ref{kasteleyn1} and \ref{kasteleyn2} below.

\begin{figure}[h]
\begin{minipage}{0.45\textwidth}
  \centering
 \includegraphics[width=0.6\textwidth]{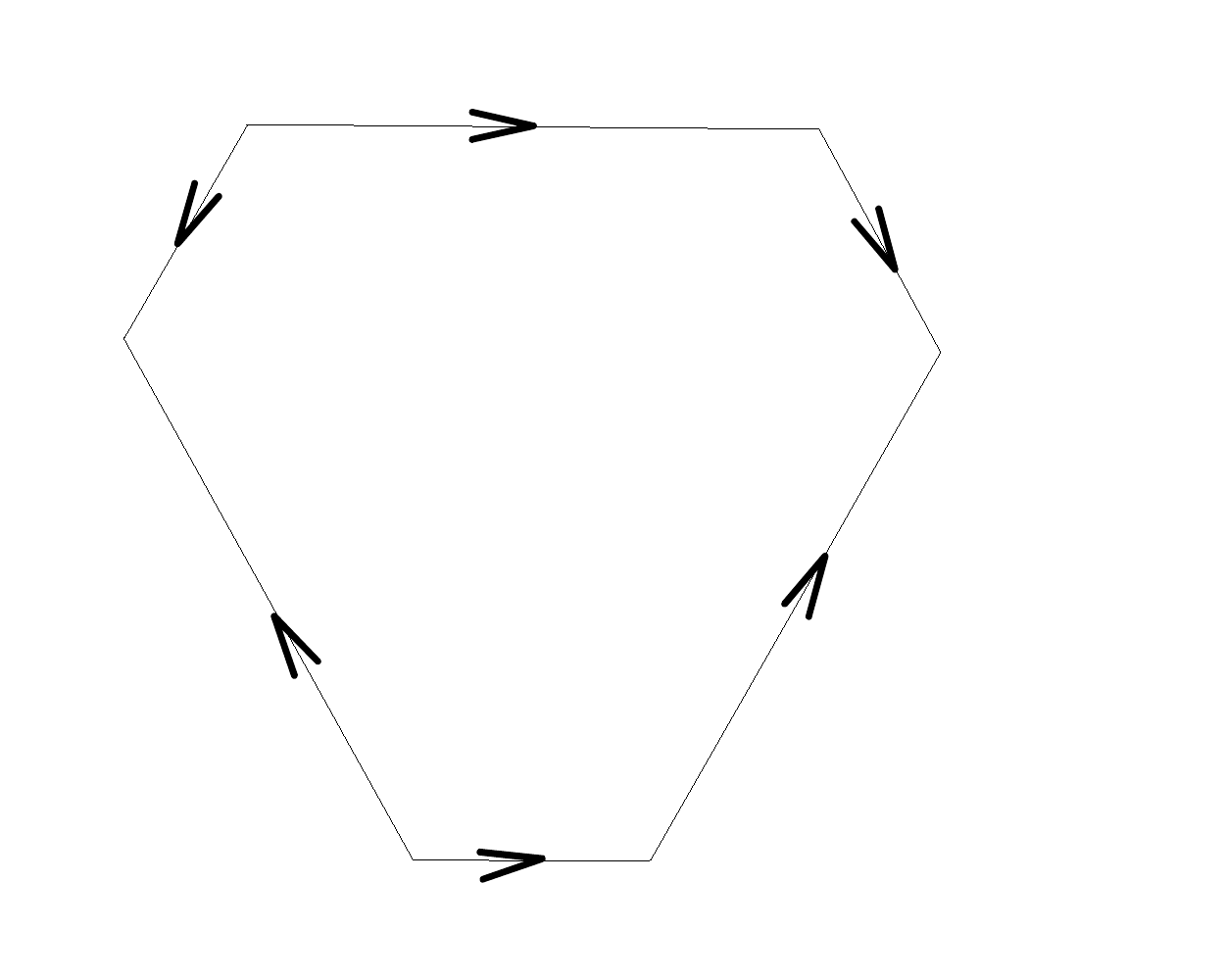}
\caption{A hexagon with Kasteleyn 
orienations.}
\label{kasteleyn1}
\end{minipage}$\qquad$
\begin{minipage}{0.45\textwidth}
\centering
 \includegraphics[width=0.6\textwidth]{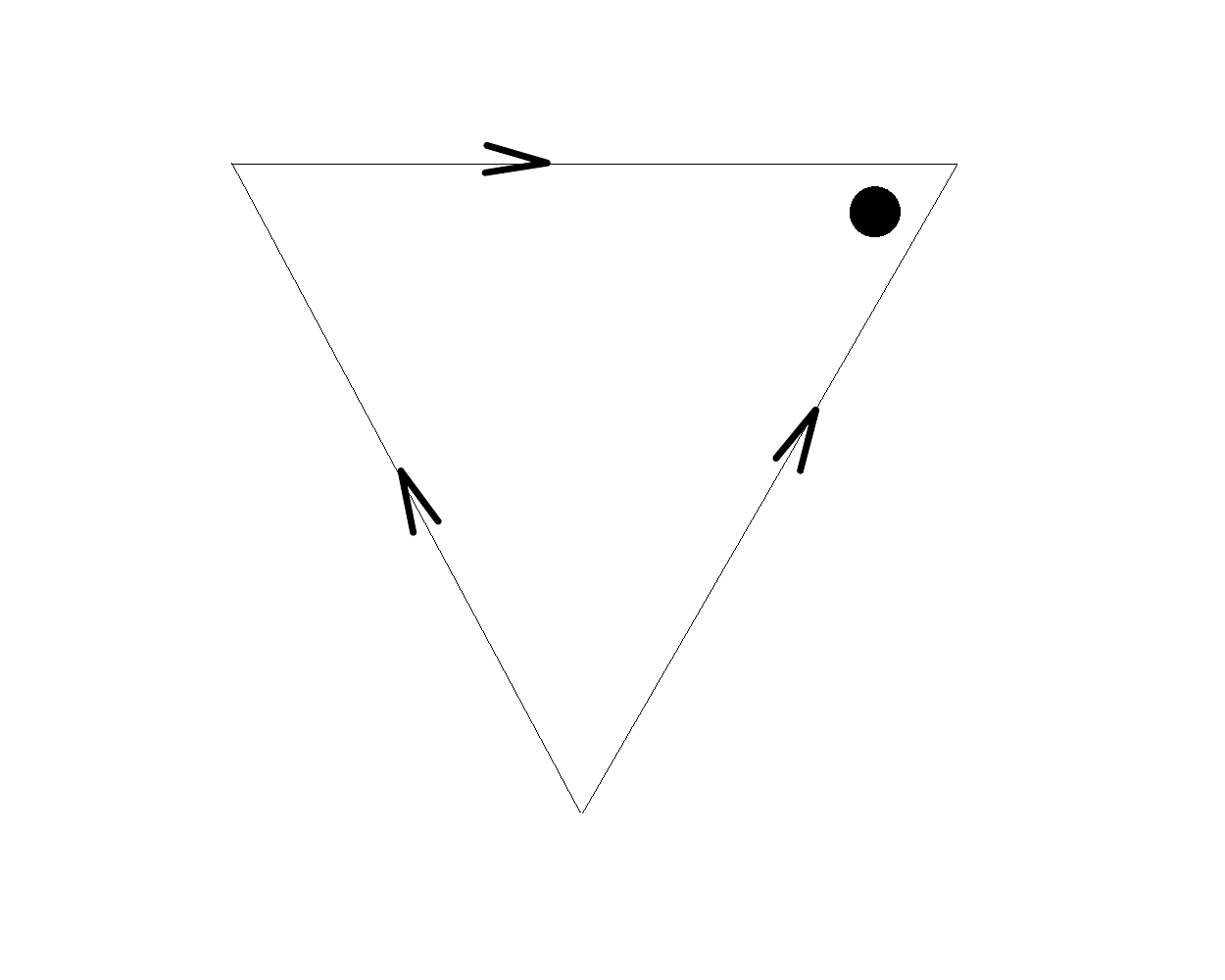}
\caption{A representation of the figure \ref{kasteleyn1} by a dotted triangle.}
\label{kasteleyn2}
\end{minipage}
\end{figure}
This amounts to representing the data encoded in a hexagonalisation with Kasteleyn orientations
in a triangulation carrying an additional decoration given by the choice of orientations for the edges, and 
by marking some corners with dots.  The data graphically 
represented by dotted triangulations will be referred to as oriented
hexagonalisations in the rest of this paper.

\subsection{Coordinates for the super \TM spaces}

In order to define  coordinates for the super \TM spaces let us consider  super Riemann surfaces
$\Sigma_{g,n} \equiv \mathbb{H}^{1|1}\slash \Gamma$ with $n\geq 1$ punctures. 
$\Sigma_{g,n}$ can be represented by
a polygonal fundamental domain $D\subset\mathbb{H}^{1|1}$ with a boundary represented by 
a collection of arcs pairwise identified with each other by the elements of $\Gamma$.
The corners of the fundamental domains $P_{i}=(x_i|\theta_i)$,
$i=1,\dots,d$ of $D$ will be located 
on the boundary $\mathbb{P}^{1|1}$ of $\mathbb{H}^{1|1}$. An ideal triangulation of the underlying 
Riemann surface $\Sigma_{g,n}^\sharp$ induces a triangulation of the 
super Riemann surface with vertices represented by the corners $P_{i}=(x_i|\theta_i)$,
$i=1,\dots,d$.
Following \cite{BB} we will in the following assign even coordinates to the 
edges of a dotted triangulation, and odd variables to the triangles themselves. 

In order to define the coordinates associated to edges let us assume that the edge $e$ represents the diagonal
in a quadrangle with corners at $P_i=(x_i |\theta_i)\in\mathbb{P}^{1|1}$, $i=1,\dots,4$ connecting $P_2$ and 
$P_4$. One may then define the even variable $z_e$ assigned to the edge $e$ to be given by the 
even superconformal cross-ratio defined in  \eqref{conformalinvariantt}.

\begin{figure}[h]
\centering
\includegraphics[width=0.3\textwidth]{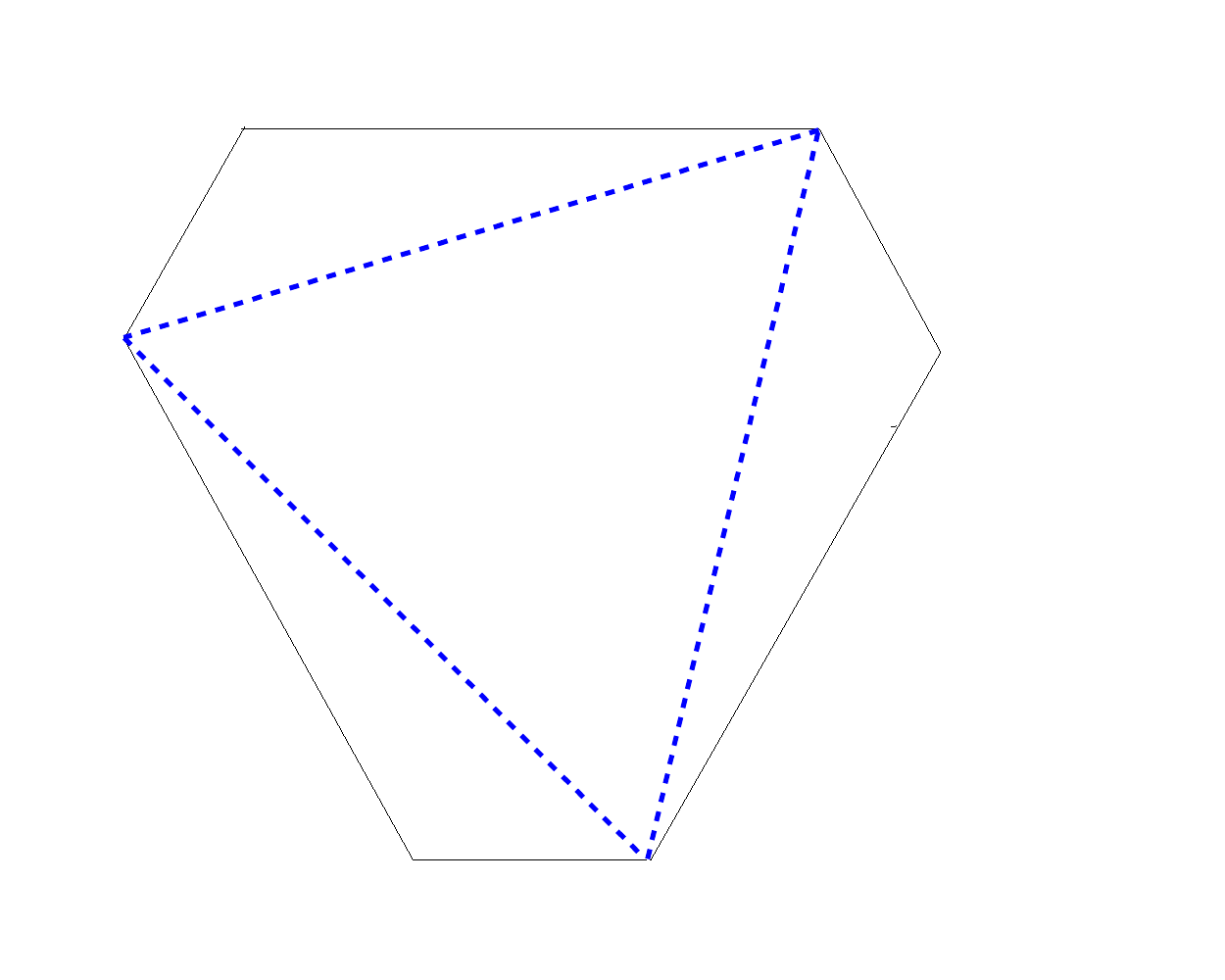}
\caption{A hexagon and its underlying triangle.}
\label{hexagon}
\end{figure}

In order to define the odd Fock variables let us consider a hexagonalisation decorated with a 
Kasteleyn orientation. We may triangulate each hexagon as
shown in Figure \ref{hexagon}. 

Note that the orientation on the sides of the hexagon induces a
canonical Kasteleyn orientation on each of the triangles appearing in this
triangulation of the hexagon.
We may therefore apply the definition of odd invariant given in Section \ref{invariants} 
to the corners of the inner triangle drawn with blue, dashed sides in Figure \ref{hexagon}.
As the hexagons of the considered hexagonalisation are in one-to-one correspondence
with the triangles $\Delta$ of a dotted triangulation we will denote the resulting 
coordinates by $\xi_\Delta$.

The super \TM space is thereby parametrised by $3(2g-2+n)$ even coordinates and $2(2g-2+n)$ odd coordinates. It 
has a super Poisson structure\footnote{A super Poisson algebra is a super-algebra $A$ with grading of $x\in A$ 
denoted as $|x|$,  which has a super Poisson bracket $\{.,.\}:A\times A\rightarrow A$ that is 
graded skew-symmetric, $\{x,y\}=-(-1)^{|x||y|}\{y,x\}$, and satisfies  
$\{x,\{y,z\}\}+(-1)^{|x|(|y|+|z|)}\{y,\{z,x\}\}+(-1)^{|z|(|x|+|y|}\{z,\{x,y\}\}$
along with $\{x,yz\}=\{x,y\}z+(-1)^{|x||y|}y\{x,z\}$.}
\cite{BB} with non-trivial  Poisson brackets among the
coordinate functions being  
\begin{equation}
\{ z_e,z_f\}_{\rm ST}^{}=n_{ef}\,,\qquad \{\xi_v,\xi_{w}\}_{\rm ST}^{}=\frac{1}{2}\delta_{vw}\,.
\label{superpoisson}
\end{equation}
where  the numbers $n_{ef}$ are defined in the same way as in ordinary \TM theory. This defines the Poisson-structure we aim to quantise.

\subsection{Super Ptolemy groupoid}

The coordinates that we use to parametrise the super \TM space depend on the choice of the dotted triangulation. It is therefore necessary to determine how those coordinates transform under the moves that change the dotted 
triangulations of the Riemann surfaces. In addition to the supersymmetric analogue of the flip operation changing 
the diagonal in a quadrilateral we now need to consider an additional move describing a change of Kasteleyn orientation which leaves the spin structure unchanged. The groupoid generated by the  changes of dotted 
triangulations will be called super Ptolemy groupoid. We will now offer a description in terms of generators and
relations.

\subsubsection{Generators}

\begin{figure}[h]
\centering
\includegraphics[width=0.5\textwidth]{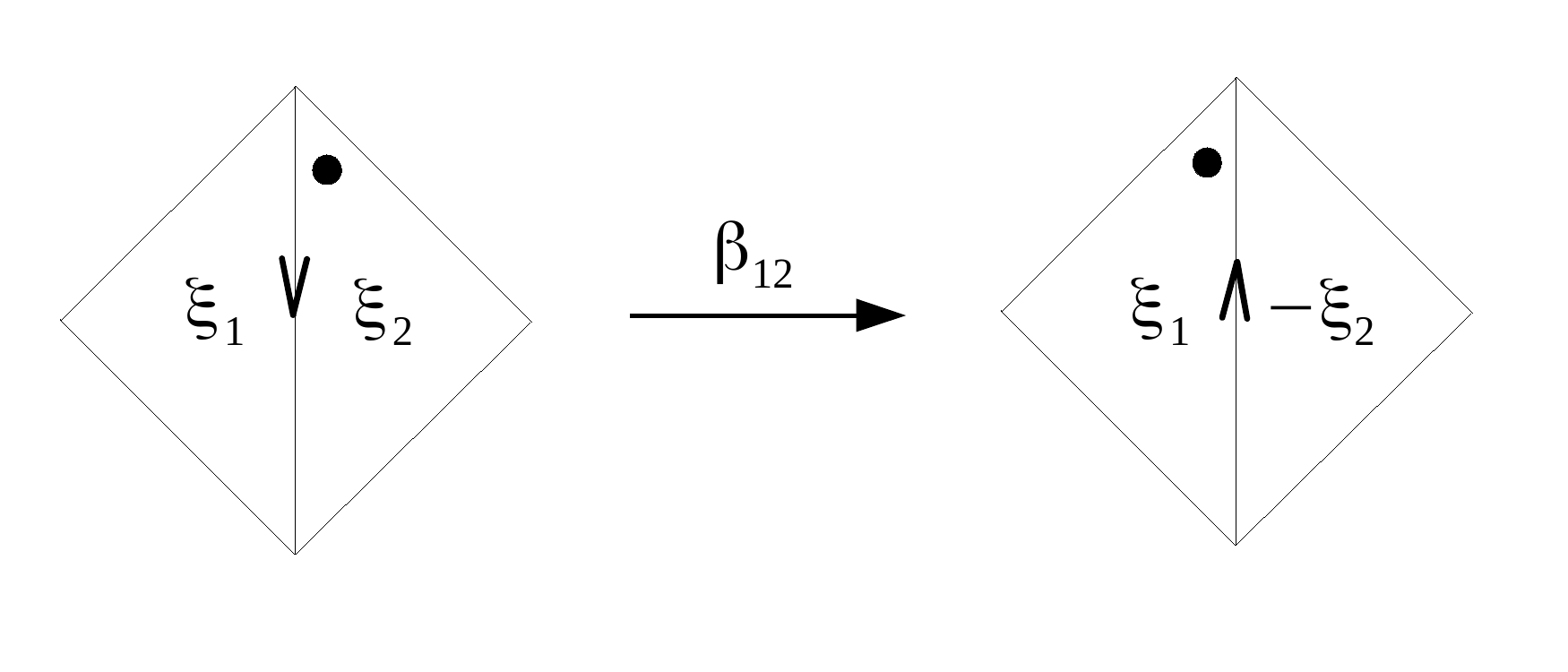}
\caption{The pictorial representation of a (left) push-out on triangles with one dot.}
\label{pushleftdot}
\end{figure}

As we discussed previously, the reversal of Kasteleyn orientations of all the edges that meet in the same vertex does not change the spin structure. Therefore, we can consider a pair of two hexagons that meet along one long edge, and study a move that applies this operation at one of the vertices common to both hexagons.
In terms of dotted triangles, one can pictorially represent this move as in the figure \ref{pushleftdot}.
 We will call this move a (left) push-out $\beta$. As for the action on the odd invariants, a push-out leaves the one of the left hexagon unchanged, but it changes the sign for the one on the right, and it does not change any of even invariants.

\begin{figure}[h]
\centering
\includegraphics[width=0.7\textwidth]{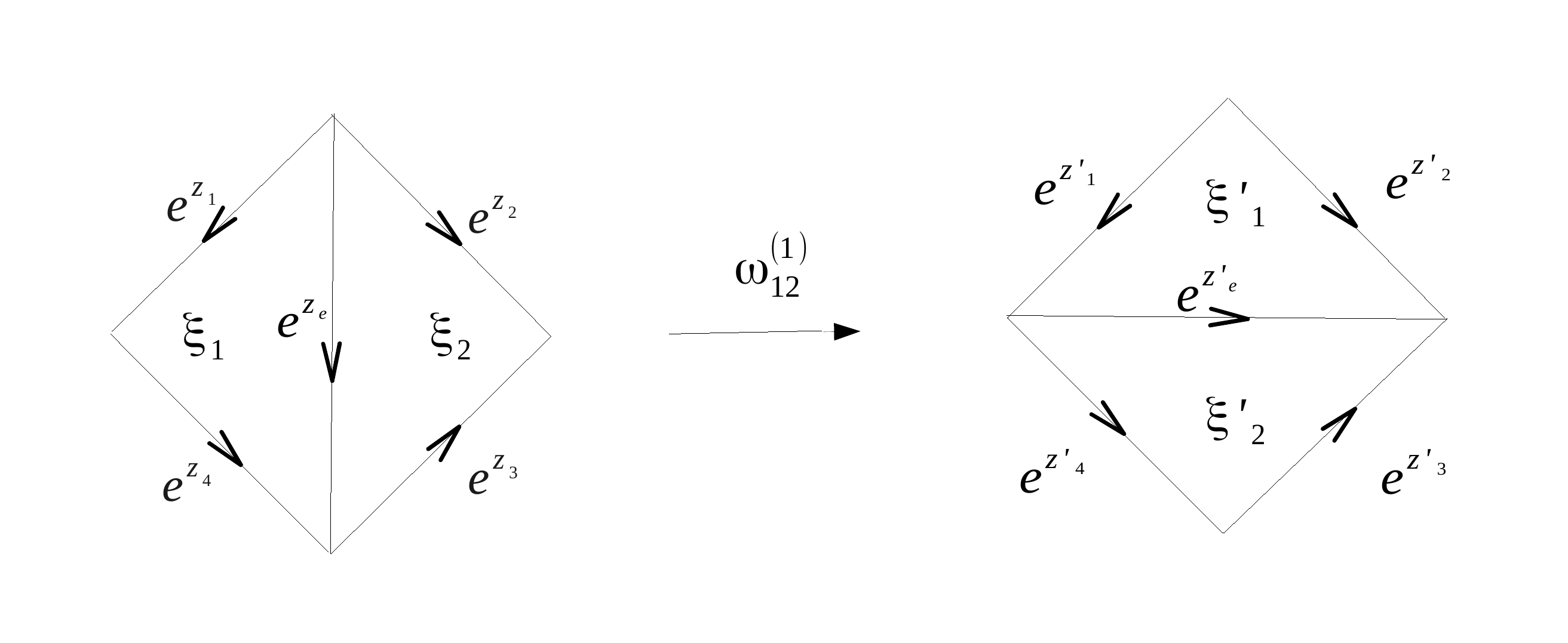}
\caption{The flip $\omega^{(1)}_{12}$.}
\label{superflipT1}
\end{figure}

We furthermore need to consider the flip operation describing the change of diagonal in a quadrilateral. 
The effect of this operation will in general depend on the assignment of Kasteleyn orientations.
An example is depicted in Figure \ref{superflipT1}.

\begin{figure}[h]
\centering
\includegraphics[width=0.5\textwidth]{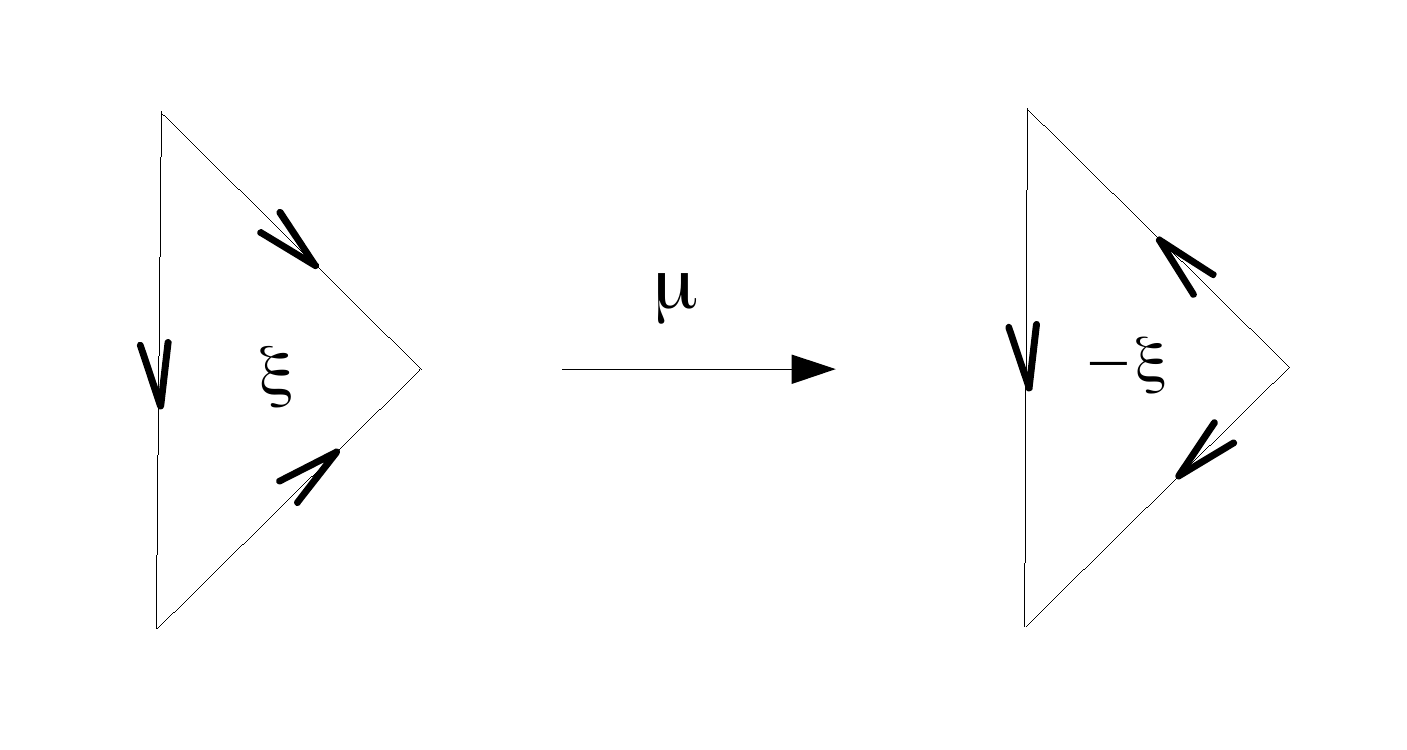}
\caption{The operation $\mu$.}\label{operatorMM}
\end{figure}

The change of even Fock coordinates may be represented as \cite{BB}
\begin{equation}\label{superflipclass1}\begin{split}
 &e^{z'_e} = e^{-z_e},\\
 &e^{z'_1} = e^{\frac{z_1}{2}}(1 + e^{z_e} - \xi_1\xi_2 e^{\frac{z_e}{2}})e^{\frac{z_1}{2}},\\
 &e^{z'_2} = e^{\frac{z_2}{2}}(1 + e^{-z_e} - \xi_1\xi_2 e^{-\frac{z_e}{2}})^{-1}e^{\frac{z_2}{2}},\\
 &e^{z'_3} = e^{\frac{z_3}{2}}(1 + e^{z_e} - \xi_1\xi_2 e^{\frac{z_e}{2}})e^{\frac{z_3}{2}},\\
 &e^{z'_4} = e^{\frac{z_4}{2}}(1 + e^{-z_e} - \xi_1\xi_2 e^{-\frac{z_e}{2}})^{-1}e^{\frac{z_4}{2}},
 \end{split}
\end{equation}
To reduce the number of 
cases to be considered in the statement of the 
transformation of the odd coordinates one may first note that the push-out operation allows one to reduce the 
most general case to the case of undotted triangles. 
It is easy to convince oneself that there are 8 possible ways of assigning Kasteleyn orientations in this case,
represented by Figure \ref{superflips1}  in the Appendix \ref{appendix2}. 
Let us begin by considering the operation $\omega^{(1)}$ depicted in Figure \ref{superflipT1}.
One then finds  the following change of coordinates \cite{BB}
\begin{equation}\label{superflipclass2}\begin{split}
 &e^{\frac{z'_1}{2}}\xi'_1 = e^{\frac{z_1}{2}}(\xi_1+\xi_2 e^{\frac{z_e}{2}}),\\
 &e^{\frac{z'_1}{2}}\xi'_2 = e^{\frac{z_1}{2}}(-\xi_1 e^{\frac{z_e}{2}} +\xi_2 ).
 \end{split}
\end{equation}
As a useful book-keeping device for generating the expressions in the other cases let us 
introduce an operation $\mu_v$ that  reverses the orientations of the two long edges entering a common vertex
of a dotted triangulation. 
This operation is graphically represented in Figure \ref{operatorMM}.
It is easy to see that this will induce a sign change in the definition of the odd invariant.

\begin{figure}[h]
\centering
\includegraphics[width=0.5\textwidth]{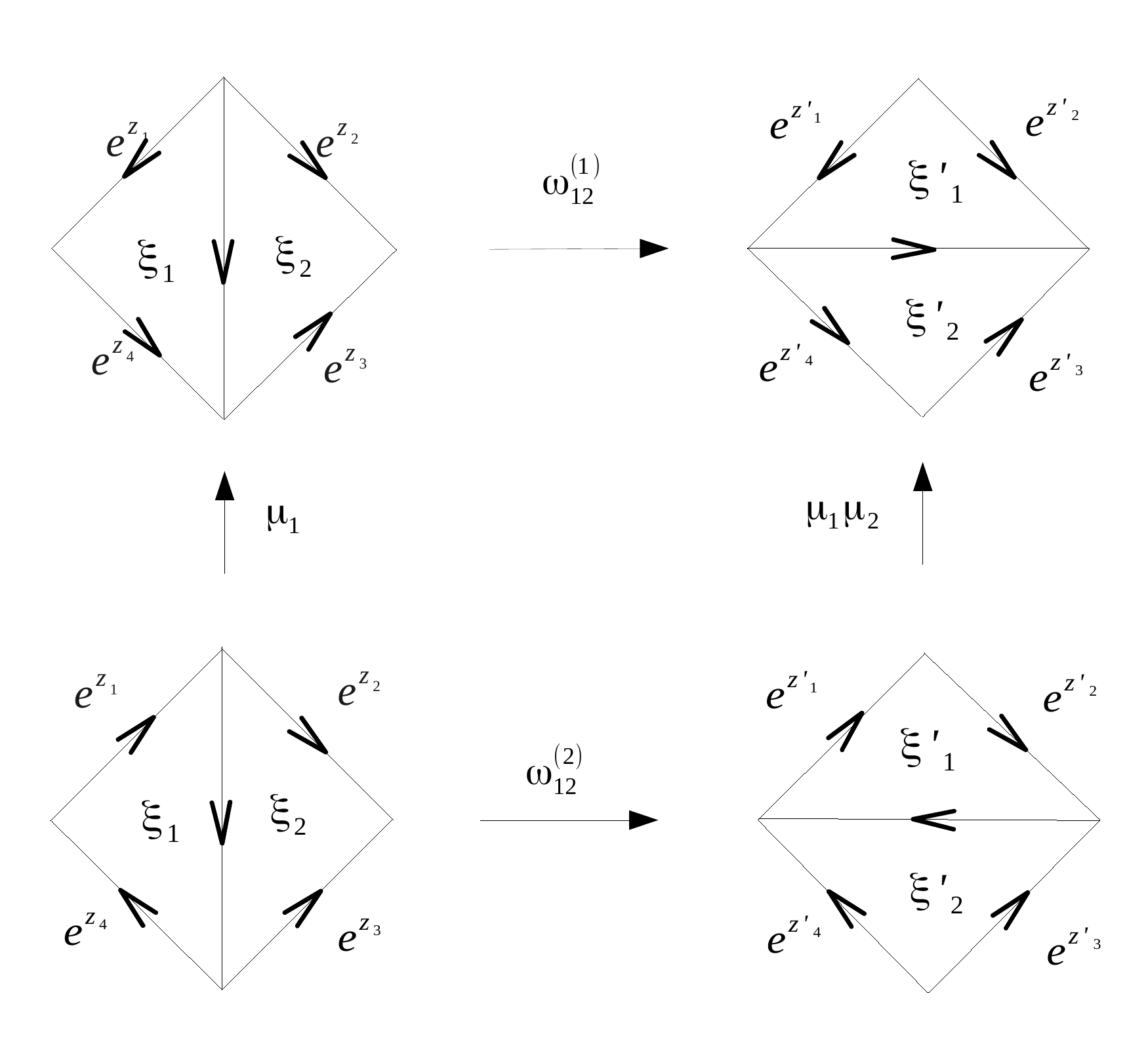}
\caption{Different flips are related by application of transformations $\mu$.}
\label{exampleflips}
\end{figure}

The coordinate transformations induced by flips with other assignments of Kasteleyn orientations can 
then be obtained from the case of $\omega^{(1)}$ with the help of the operations $\mu_v$. An example is 
represented by Figure \ref{exampleflips}.

\subsubsection{Relations}

The  changes of oriented hexagonalisations define a groupoid generalising the Ptolemy groupoid. In the 
following we are going to discuss the relations characterising this groupoid which will be called super 
Ptolemy groupoid.

It is clear that all relations of the super Ptolemy groupoid reduce to relations of the ordinary
Ptolemy groupoid upon forgetting the decorations furnished by the Kasteleyn orientations.
This fact naturally allows us to distinguish a few different types of relations. 

\begin{figure}[!h]
\centering
 \includegraphics[width=0.6\textwidth]{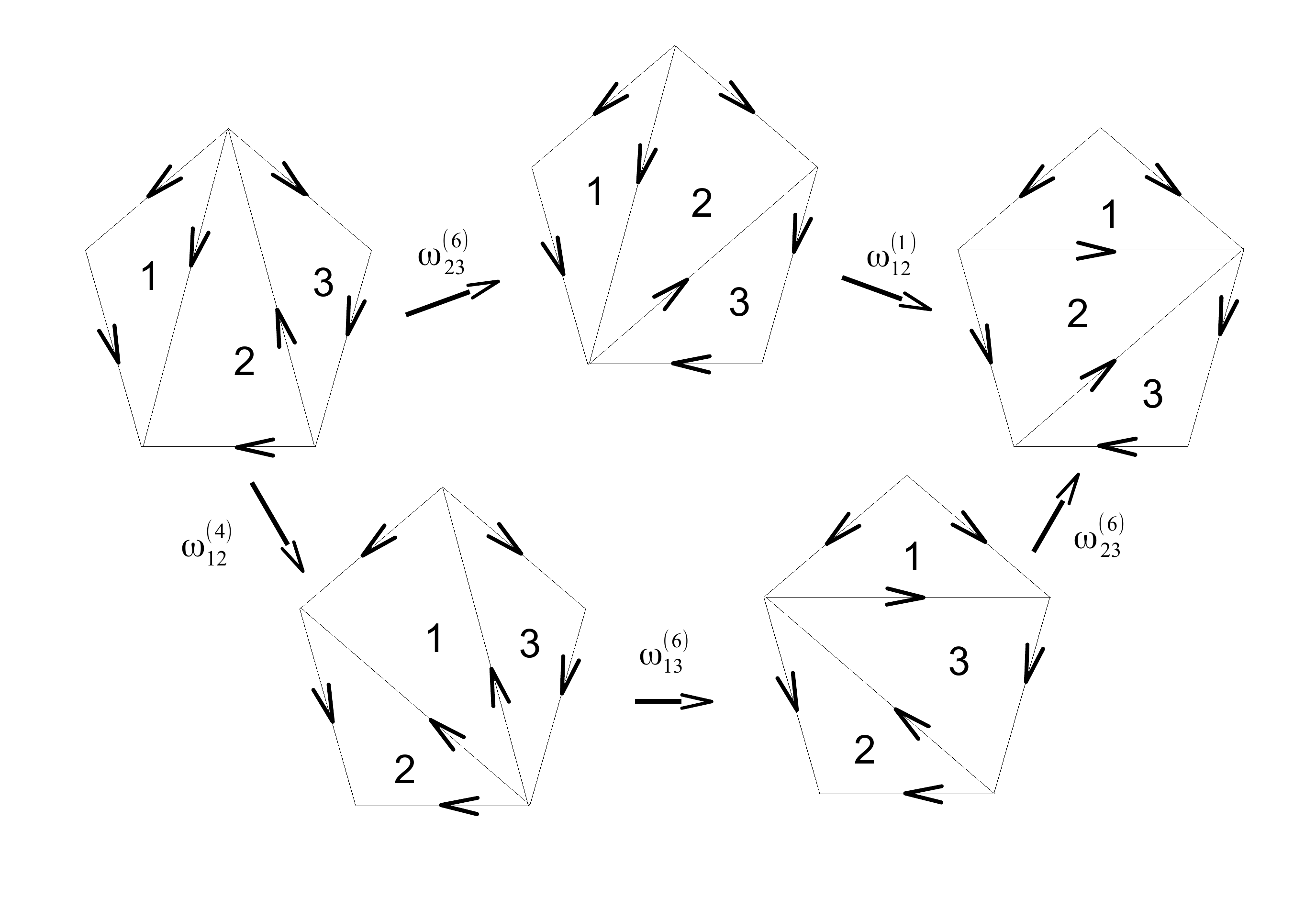}
\caption{An example of one of the possible superpentagon relations. }
\label{superpentagonclassicexample}
\end{figure}

To begin with, let us consider the relations reducing to the pentagon relation of the Ptolemy groupoid.
The super Ptolemy groupoid will have various relations differing by the  choices of Kasteleyn orientations. 
Considering first the case where all short edges are oriented with the orientation of the surface
we have 16 possible pentagon relations:
\begin{equation}\begin{split}
&\omega^{(1)}_{12} \omega^{(1)}_{23}= \omega^{(1)}_{23} \omega^{(1)}_{13} \omega^{(1)}_{12}, \qquad \omega^{(1)}_{12} \omega^{(6)}_{23}= \omega^{(6)}_{23} \omega^{(6)}_{13} \omega^{(4)}_{12} ,\\
&\omega^{(5)}_{12} \omega^{(8)}_{23}= \omega^{(8)}_{23} \omega^{(5)}_{13} \omega^{(5)}_{12}, \qquad \omega^{(6)}_{12} \omega^{(7)}_{23} = \omega^{(7)}_{23} \omega^{(6)}_{13} \omega^{(5)}_{12},\\
& \omega^{(2)}_{12} \omega^{(1)}_{23}= \omega^{(1)}_{23} \omega^{(2)}_{13} \omega^{(2)}_{12}, \qquad \omega^{(8)}_{12} \omega^{(8)}_{23} = \omega^{(1)}_{23} \omega^{(8)}_{13} \omega^{(8)}_{12},\\
& \omega^{(4)}_{12} \omega^{(5)}_{23}= \omega^{(5)}_{23} \omega^{(5)}_{13} \omega^{(4)}_{12}, \qquad \omega^{(5)}_{12} \omega^{(3)}_{23} = \omega^{(3)}_{23} \omega^{(4)}_{13} \omega^{(6)}_{12},\\
& \omega^{(3)}_{12} \omega^{(4)}_{23}= \omega^{(7)}_{23} \omega^{(3)}_{13} \omega^{(2)}_{12}, \qquad \omega^{(7)}_{12} \omega^{(7)}_{23} = \omega^{(4)}_{23} \omega^{(7)}_{13} \omega^{(8)}_{12},\\
& \omega^{(6)}_{12} \omega^{(2)}_{23} = \omega^{(2)}_{23} \omega^{(1)}_{13} \omega^{(6)}_{12}, \qquad \omega^{(7)}_{12} \omega^{(2)}_{23} = \omega^{(5)}_{23} \omega^{(2)}_{13} \omega^{(7)}_{12},\\
& \omega^{(5)}_{12} \omega^{(6)}_{23}= \omega^{(3)}_{23} \omega^{(7)}_{13} \omega^{(6)}_{12}, \qquad \omega^{(3)}_{12} \omega^{(5)}_{23} = \omega^{(2)}_{23} \omega^{(8)}_{13} \omega^{(3)}_{12},\\
& \omega^{(1)}_{12} \omega^{(3)}_{23}= \omega^{(6)}_{23} \omega^{(3)}_{13} \omega^{(7)}_{12}, \qquad \omega^{(4)}_{12} \omega^{(4)}_{23} = \omega^{(4)}_{23} \omega^{(4)}_{13} \omega^{(1)}_{12}.
\end{split}\end{equation}
The remaining cases can always be reduced to the cases listed above using the push-out operation.
In Figure \ref{superpentagonclassicexample} we present one of the 16 possibilities listed above graphically.

\begin{figure}[h] 
\centering
\includegraphics[width=0.6\textwidth]{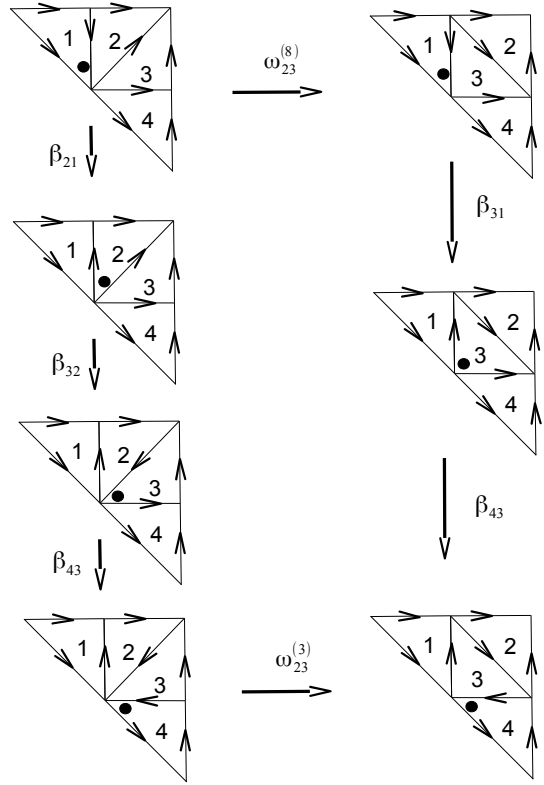}
\caption{First type of relation between a flip and a push-out.} \label{dot1example}
\end{figure}

Other relations reduce to trivial  relations upon forgetting the orientation data. Some of these relations
describe how the push-out operations  relate flips with different orientation data.
Such relations are 
\begin{equation}
 (\omega^{(i)}_{23})^{-1} \beta_{43}^{} \beta_{32}^{}
\beta_{21}^{}= \beta_{42}^{} \beta_{21}^{}(\omega^{(j)}_{23})^{-1},
\end{equation}
where $i,j$ can be following pairs $(5,8),(8,5),(6,7),(7,6),(1,2),(2,1),(3,4),(4,3)$ and
\begin{equation}
 \omega^{(i)}_{23} \beta_{43}^{} \beta_{32}^{} \beta_{21}^{} = \beta_{43}^{} \beta_{31}^{} \omega^{(j)}_{23},
\end{equation}
where $i,j$ can be following pairs $(5,4),(4,5),(1,6),(6,1),(2,7),(7,2),(8,3),(3,8)$. An example for this type of relation
is illustrated in Figure \ref{dot1example}. 

\begin{figure}[h] 
\centering
\includegraphics[width=0.6\textwidth]{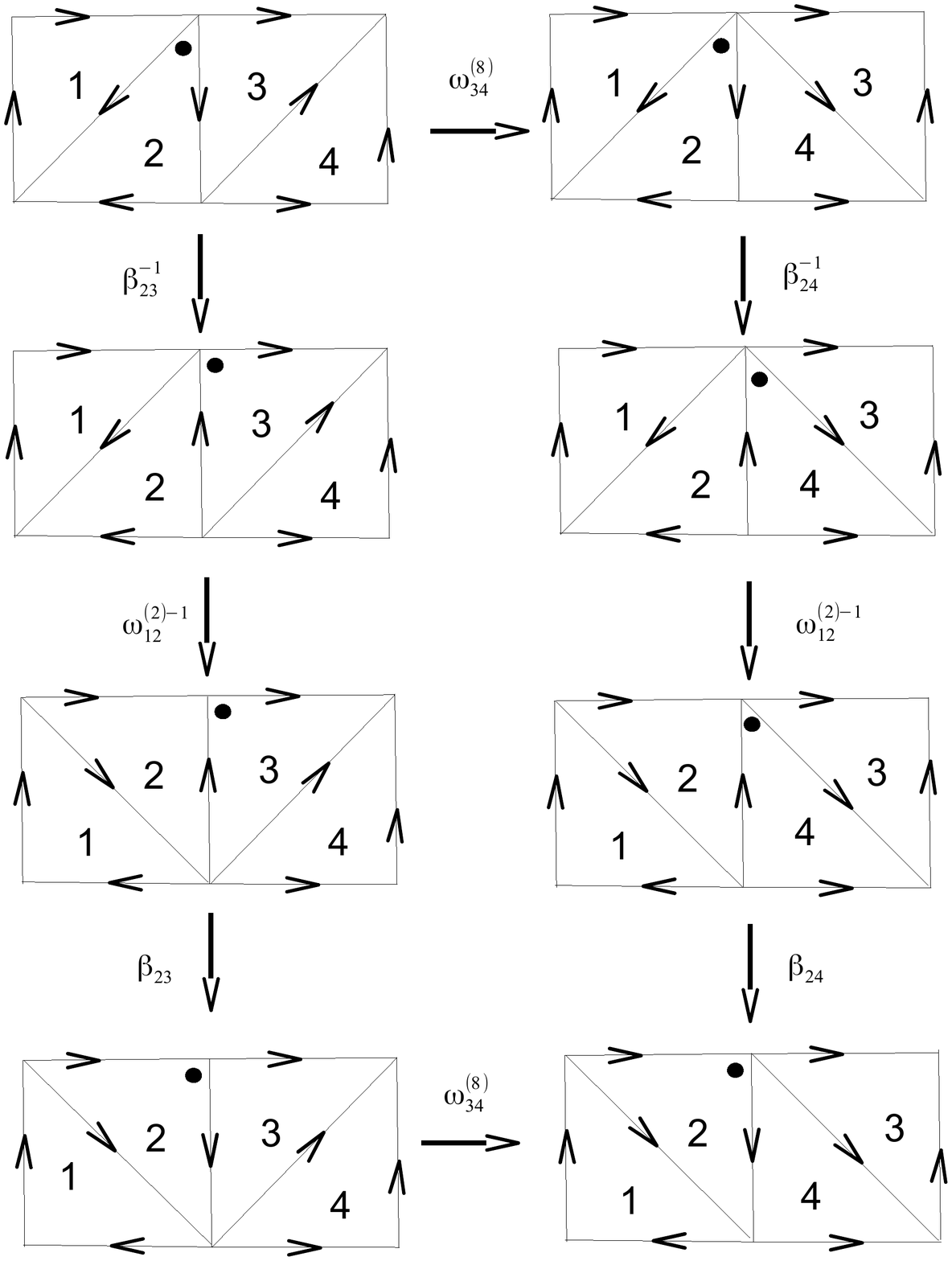}
\caption{Second type of relation between a flip and a push-out.}\label{dot2example}
\end{figure}

There are further relations reducing to the commutativity of the flip operations applied to two quadrilaterals which
do not share a triangle, including
\begin{align}
&\omega_{34}^{(i)} \beta_{23}^{} (\omega_{12}^{(j)})^{-1} \beta_{23}^{-1} = \beta_{24}^{} (\omega_{12}^{(j)})^{-1} \beta_{24}^{-1} \omega_{34}^{(i)}, \\
&(\omega_{34}^{(i)})^{-1} \beta_{13}^{} \omega_{12}^{(j)} \beta_{23}^{-1} = \beta_{13}^{} \omega_{12}^{(j)} \beta_{23}^{-1} (\omega_{34}^{(i)})^{-1},\\
&(\omega_{34}^{(i)})^{-1} \beta_{23}^{} (\omega_{12}^{(j)})^{-1} \beta_{23}^{-1} = \beta_{23}^{} (\omega_{12}^{(j)})^{-1} \beta_{23}^{-1} (\omega_{34}^{(i)})^{-1},\\
&\omega_{34}^{(i)} \beta_{13}^{} \omega_{12}^{(j)} \beta_{23}^{-1} = \beta_{14}^{} \omega_{12}^{(j)} \beta_{24}^{-1} \omega_{34}^{(i)}, 
\end{align}
where the $i,j=1,\ldots,8$ depends on the Kasteleyn orientation of the graph from which the relation has been derived. 
Examples of these relations are represented  in Figure \ref{dot2example}.

It seems plausible that the completeness of the relations discussed above can be reduced to  
the corresponding result for the ordinary Ptolemy groupoid. This result, as pointed out in \cite{Chekhov:1999tn},
follows from the cell decomposition of the Teichm\"uller space which can be defined with the help of
Penner's coordinates \cite{Penner}.

\subsection{Kashaev type coordinates}

It will furthermore be useful to introduce analogues of the 
Kashaev coordinates in the case of super \TM theory. 
Such coordinates will be associated to oriented hexagonalisations carrying an additional piece of 
decoration obtained by marking a distinguished short edge in each hexagon. Oriented hexagonalisations
equipped with such a decoration will be called decorated hexagonalisations in the following.

In addition to a pair of even variables $(q_v,p_v)$ assigned to each ideal triangle $\Delta_v$, we now need to 
introduce an odd variable $\xi_v$. 
The collection of these variables parametrising points in
 $\mathbb{R}^{4(2g-2+n)|2(2g-2+n)}$, which we will name super Kashaev space, will be called super Kashaev coordinates. 
 The non-trivial Poisson brackets defining the Poisson structure on this space are 
\begin{equation}
 \{p_v,q_w\}_{\rm ST}^{} = \delta_{v,w},\qquad
    \{\xi_v,\xi_w\}_{\rm ST}^{} = \frac{1}{2}\delta_{v,w},
\label{kashaevpoissonbracketsusy} 
\end{equation}
all other Poisson brackets among the variables $(q_v,p_v,\xi_v)$ being trivial.

The super Teichm\"uller spaces can be characterised within 
$\mathbb{R}^{8g-8+4n|4g-4+2n}$ by using the Hamiltonian reduction with respect to a set of 
constraints that is very similar to the one used in ordinary Teichm\"uller theory
described in \cite{Kash1}. 
One may, in particular, recover the even Fock coordinates 
in a way that is very similar to \eqref{ze}, while the odd variables simply coincide.

\begin{figure}[h] 
\centering
\includegraphics[width=0.7\textwidth]{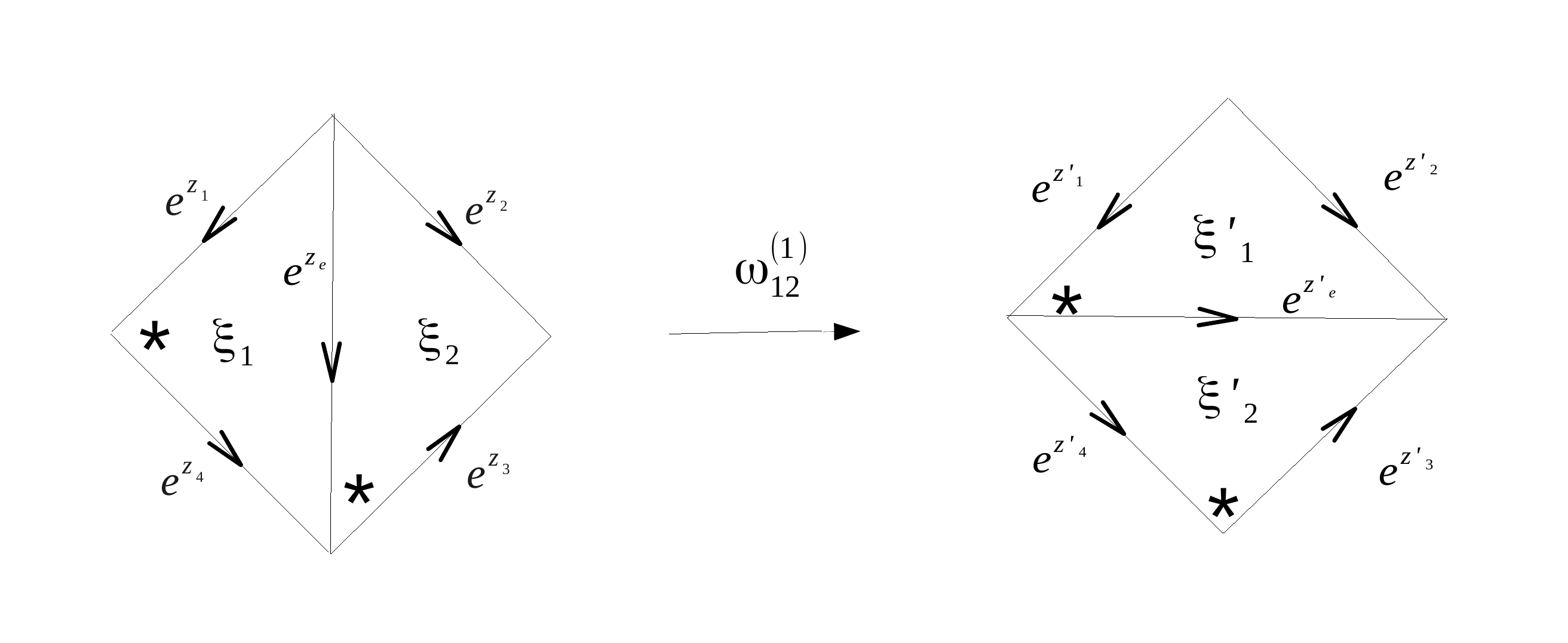}
\caption{A flip $\omega^{(1)}$ on decorated triangulation.  }\label{flipkashaevclass}
\end{figure}

The transformations relating different decorated hexagonalisations will induce 
changes of super Kashaev coordinates. Such transformations will generate a decorated version of 
the super Ptolemy groupoid. The set of generators becomes as in the case of ordinary
Teichm\"uller theory enriched by the 
operation $(vw)$ exchanging the labels associated to two adjacent triangles, 
and the rotations $\rho_v$ of the distinguished short edge.
The rotation $\rho_v$ will be represented as
\begin{equation}
 \rho_v^{-1}:(q_v,p_v,\xi_v)\rightarrow(p_v-q_v,-q_v,\xi_v).
\end{equation}
The operation $(vw)$ maps $(q_v,p_v,\xi_v)$ to $(q_w,p_w,\xi_w)$ and vice-versa.
The flip $\omega_{vw}^{(1)}$, presented in the figure \ref{flipkashaevclass}, is realised by 
\begin{equation}\label{kashaevflipclassicalsusy}
(\omega^{(1)}_{vw})^{-1}: \begin{cases}     
    (U_v,V_v)\rightarrow(U_vU_w,U_vV_w+V_v-U_v^\frac{1}{2}V_w^\frac{1}{2}V_v^\frac{1}{2}\xi_v\xi_w),\\
   (U_w,V_w)\rightarrow(U_wV_v(U_vV_w+V_v-U_v^\frac{1}{2}V_w^\frac{1}{2}V_v^\frac{1}{2}\xi_v\xi_w)^{-1},\\
   \qquad V_w(U_vV_w+V_v-U_v^\frac{1}{2}V_w^\frac{1}{2}V_v^\frac{1}{2}\xi_v\xi_w)^{-1}),
\end{cases}
\end{equation}
for the even variables and
\begin{equation}\label{kashaevflipclassicalsusyodd}
(\omega^{(1)}_{vw})^{-1}: \begin{cases}     
    \xi_v \rightarrow \frac{V_v^\frac{1}{2} \xi_v + U_v^\frac{1}{2}V_w^\frac{1}{2} \xi_w}{\sqrt{V_v + U_v V_w - U_v^\frac{1}{2}V_w^\frac{1}{2}V_v^\frac{1}{2}\xi_v\xi_w}} ,\\
   \xi_w \rightarrow \frac{V_v^\frac{1}{2} \xi_w - U_v^\frac{1}{2}V_w^\frac{1}{2} \xi_v}{\sqrt{V_v + U_v V_w - U_v^\frac{1}{2}V_w^\frac{1}{2}V_v^\frac{1}{2}\xi_v\xi_w}} ,
\end{cases}
\end{equation}
for the odd ones, where we denote $U_v\equiv e^{q_v}$ and $V_v \equiv e^{p_v}$. The action of the rest of flips \footnote{The flips transforming Kashaev coordinates relate decorated versions of quadrilaterals. Therefore, to represent flips of Kashaev coordinates one should add decoration to all the figures in \ref{superflips1} in the same places as in the figure \ref{flipkashaevclass}.} can be obtained by the application of appropriate operations $\mu_v$, as explained previously . 


\section{Quantisation of super Teichm\"uller theory}\label{chapter5}
 
In this section we will consider the quantisation of the Teichm\"uller spaces of super Riemann surfaces. The coordinate functions defined in the previous section will become linear operators acting on a Hilbert space. The transformations which relate different hexagonalisations, like flips and push-outs, will be represented by linear operators $\mathsf{T}$ and $\mathsf{B}$, 
respectively. We are going to discuss the relations satisfied by these operators, defining a projective representation of 
the super Ptolemy groupoid.

\subsection{Quantisation of super Kashaev space}

The Hilbert space associated to a decorated hexagonalisation of a super Riemann surface
will be defined as follows. To each hexagon $\Delta_v$ (or equivalently each dotted triangle) we associate a Hilbert space $\mathcal{H}_v \simeq L^2(\mathbb{R})\otimes\mathbb{C}^{1|1}$. Then, the Hilbert space associated to the entire super Riemann surface is the tensor product of the spaces for each hexagon:
\begin{equation}
 \mathcal{H} = \bigotimes_{v\in I} \mathcal{H}_v .
\end{equation}
We will frequently use the corresponding leg-numbering notation: If $\mathsf{O}$ is an operator on 
$L^2(\mathbb{R})\otimes\mathbb{C}^{1|1}$, we may define $\mathsf{O}_v$ to be the operator 
$
\mathsf{O}_v=1\otimes\cdots\otimes1\otimes \underset{v-\rm th}{\mathsf{O}}\otimes 1\otimes\cdots \otimes 1\,.
$

The super Kashaev coordinates get quantised to linear operators on the Hilbert spaces $\mathcal{H}_v$. 
The coordinates $\mathsf{p}_v$ and $\mathsf{q}_v$ are replaced by operators satisfying canonical commutation relations 
 \begin{equation}
   \left[\mathsf{p}_v,\mathsf{q}_w\right]=\frac{1}{ \pi i}\delta_{vw}, \qquad
   \left[\mathsf{q}_v,\mathsf{q}_w\right]= 0, \qquad
   \left[\mathsf{p}_v,\mathsf{p}_w\right]=0 ,
 \end{equation}
 and are represented on $L^2(\mathbb{R})$ as multiplication and differentiation operators. In the classical limit $b\to0$, the operators $2\pi b\mathsf{p}$ and $2\pi b\mathsf{q}$ give their classical counterparts $p$ and $q$ appropriately. The odd coordinate $\xi$ becomes an operator acting on $\mathcal{H}$ of the form
\begin{equation}
 \xi = \sqrt{ q^\frac{1}{2}-q^{-\frac{1}{2}} } \kappa,
\end{equation}
where $\kappa$ is a $(1|1)\times(1|1)$ matrix acting on $\mathbb{C}^{1|1}$
\begin{equation}
 \kappa = \bigg( \begin{array}{cc} 0 & 1 \\ 1 & 0 \end{array} \bigg),
\end{equation}
and where $q=e^{i\pi b^2}$ and the quantisation constant $\hbar$ is related to $b$ as $\hbar= 4\pi b^2$. Note that $\xi$ satisfies $\xi^2=q^\frac{1}{2}-q^{-\frac{1}{2}}=i\pi b^2+\mathcal{O}(b^4)$,
thereby reproducing both the relation $\xi^2=0$ and the Poisson bracket $\{\xi,\xi\}=\frac{1}{2}$ in the classical limit $b\to 0$. 

Moreover, the formula (\ref{ze}), with the super coordinates replacing the ordinary ones, has an obvious counterpart in the quantum theory, defining self-adjoint even operators $\mathsf{z}_e$ satisfying
 \begin{equation}
 \left[  \mathsf{z}_e,\mathsf{z}_{e'} \right] = \frac{1}{\pi i} \left\lbrace z_e,z_{e'}\right\rbrace _{\rm ST}^{} .
 \end{equation} 
The operators $2\pi b\mathsf{z}_e$ give in the classical limit the even Fock coordinates $z_e$. 

The redundancy of the parametrisation in terms of Kashaev type coordinates can be described using a quantum version of the Hamiltonian reduction characterising the super Teichm\"uller spaces within $\mathbb{R}^{8g-8+4n|4g-4+2n}$. This procedure is very similar to the case of the usual  \TM theory described in \cite{Kash1,Kash2} and will therefore not be discussed explicitly here.

\subsection{Generators of the super Ptolemy groupoid}

We will now construct  a quantum realisation of 
the coordinate transformations  induced by changing the decorated hexagonalisation $\eta$ 
of a super Riemann surface $\Sigma$. The coordinate transformations will be represented by
operators $\mathsf{U}_{\eta'\eta}: \mathcal{H}_\eta \to \mathcal{H}_{\eta'}$ representing the change of the hexagonalisation $\eta$ to $\eta'$ in the following way. Let
$\{w^{\imath};\imath\in\mathcal{I}_\eta\}$ be a complete set of coordinates
defined in terms of a hexagonalisation $\eta$. If $\eta'$ is another hexagonalisation one
may in our case express the coordinates  $\{\tilde{w}^{\jmath};\jmath\in\mathcal{I}_{\eta'}\}$ associated to $\eta'$
as functions ${w}'{}^{\jmath}= 
W_{\eta'\eta}^{\jmath}(\{w^{\imath};\imath\in\mathcal{I}_{\eta}\})$ of the coordinates
$w^{\imath}$. If $\mathsf{w}^{\imath}$ and ${\mathsf{w}'}^{\jmath}$ are the operators
associated to ${w}_{\imath}$ and ${w}'{}_{\jmath}$, respectively, we are first going to define
quantised versions of the changes of coordinate functions 
$\mathsf{W}_{\eta'\eta}^{\jmath}(\{\mathsf{w}_{\imath};\imath\in\mathcal{I}_{\eta}\})$ which reduce to 
the functions $W^{\eta'\eta}_{\jmath}$ in the classical limit. Unitary operators 
$ \mathsf{U}_{\eta'\eta}^{}$ representing these changes of coordinates on the quantum level 
are then required to satisfy
\begin{equation}\label{transfgeneral}
 \mathsf{U}_{\eta'\eta}^{-1} \cdot {\mathsf{w}'}^{\jmath} \cdot \mathsf{U}_{\eta'\eta}^{} = 
 \mathsf{W}_{\eta'\eta}^{\jmath}(\{\mathsf{w}_{\imath};\imath\in\mathcal{I}_{\eta}\})\,.
\end{equation}
This requirement is expected to characterise the operators $ \mathsf{U}_{\eta'\eta}^{}$ uniquely up to 
normalisation.
We are now going to construct the operators $\mathsf{U}_{\eta'\eta}^{}$ for all pairs $\eta$ and $\eta'$
related by generators of the super Ptolemy groupoid.

Of particular interest are the cases where $\eta$ and $\eta'$ are related by the 
flip operation changing the diagonal in a triangulation.  We will begin by constructing 
operators 
$\mathsf{T}^{(i)}_{vw}:\mathcal{H}_v\otimes \mathcal{H}_w \to \mathcal{H}_v\otimes \mathcal{H}_w$, $i=1,\dots,8$ representing
the super flips of hexagonalisations listed in Appendix \ref{appendix2}, with decorated vertices placed in appropriate places. In order to cover the remaining cases one may
use the push-out operation, as will be discussed later. A useful starting point will be the 
operator $\mathsf{T}^{(1)}_{12}$ corresponding to the operation $\omega^{(1)}_{12}$ depicted in figure \ref{flipkashaevclass}. Following the discussion around (\ref{transfgeneral}) above, we will require 
that
\begin{subequations}\label{flipoperator}
\begin{equation}
 \begin{split}
  & {\mathsf{T}^{(1)}_{12}}^{-1} e^{2 \pi b \mathsf{z}'_1} {\mathsf{T}^{(1)}_{12}} = e^{ \pi b \mathsf{z}_1} (1 + e^{2 \pi b \mathsf{z}_e} - e^{ \pi b \mathsf{z}_e} \xi_1 \xi_2) e^{ \pi b \mathsf{z}_1}, \\
  & {\mathsf{T}^{(1)}_{12}}^{-1} e^{2 \pi b \mathsf{z}'_2} {\mathsf{T}^{(1)}_{12}} = e^{ \pi b \mathsf{z}_2} (1 + e^{-2 \pi b \mathsf{z}_e} - e^{- \pi b \mathsf{z}_e} \xi_1 \xi_2)^{-1} e^{ \pi b \mathsf{z}_2}, \\
  & {\mathsf{T}^{(1)}_{12}}^{-1} e^{2 \pi b \mathsf{z}'_3} {\mathsf{T}^{(1)}_{12}} = e^{ \pi b \mathsf{z}_3} (1 + e^{2 \pi b \mathsf{z}_e} - e^{ \pi b \mathsf{z}_e} \xi_1 \xi_2) e^{ \pi b \mathsf{z}_3}, \\
  & {\mathsf{T}^{(1)}_{12}}^{-1} e^{2 \pi b \mathsf{z}'_4} {\mathsf{T}^{(1)}_{12}} = e^{ \pi b \mathsf{z}_4} (1 + e^{-2 \pi b \mathsf{z}_e} - e^{- \pi b \mathsf{z}_e} \xi_1 \xi_2)^{-1} e^{ \pi b \mathsf{z}_4}, \\
  & {\mathsf{T}^{(1)}_{12}}^{-1} e^{2 \pi b \mathsf{z}'_e} {\mathsf{T}^{(1)}_{12}} = e^{-2 \pi b \mathsf{z}_e} ,
 \end{split}
\end{equation}
for the even coordinates and
%
%
\begin{equation}
 \begin{split}
 {\mathsf{T}^{(1)}_{12}}^{-1} e^{  \pi b \mathsf{z}'_1} \xione' {\mathsf{T}^{(1)}_{12}} &=  e^{\frac{1}{2 } \pi b \mathsf{z}_1} (\xione + {e^{ \pi b \mathsf{z}_e}} \xitwo) e^{\frac{1}{2 } \pi b \mathsf{z}_1}, \\
 {\mathsf{T}^{(1)}_{12}}^{-1} e^{  \pi b \mathsf{z}'_1} \xitwo' {\mathsf{T}^{(1)}_{12}} &=  e^{\frac{1}{ 2} \pi b \mathsf{z}_1} (- {e^{ \pi b \mathsf{z}_e}} \xione + \xitwo ) e^{\frac{1}{2 } \pi b \mathsf{z}_1}, \\
 \end{split}
\end{equation}
\end{subequations}
for the odd ones.  The labelling of variables is the one introduced in Figure \ref{flipkashaevclass}, 
and the definition of the variables $\mathsf{z}_e$ in terms of the Kashaev type variables uses the same conventions
as introduced in Section \ref{bos-kash} above.

An 
operator $\mathsf{T}^{(1)}_{12}$ satisfying (\ref{flipoperator}) can be constructed as follows
\begin{equation}
 \begin{aligned}
  \mathsf{T}^{(1)}_{12} &= \frac{1}{2} \Big[\,
  f_+(\mathsf{q}_1+\mathsf{p}_2-\mathsf{q}_2) 
  - i  f_-(\mathsf{q}_1+\mathsf{p}_2-\mathsf{q}_2)\,\kappa_1\,\kappa_2 \,\Big] e^{-i\pi \mathsf{p}_1 \mathsf{q}_2} .
 \end{aligned}
\end{equation}
The operator $\mathsf{T}^{(1)}_{12}$ is unitary and satisfies (\ref{flipoperator}) if
$f_{\pm}(x):=e_\R(x)\pm e_\NS(x)$ with $e_\NS(x)$ and $e_\R(x)$ being special functions satisfying 
$|e_\NS(x)|=1$ and $|e_\R(x)|=1$ for $x\in\mathbb{R}$, together with the functional relations
\begin{align*}
 e_\R\left(x-\frac{i b^{\pm1}}{2}\right) &= (1+ie^{\pi b^{\pm1}x})e_\NS\left(x+\frac{ib^{\pm1}}{2}\right), \\
 e_\NS\left(x-\frac{i b^{\pm1}}{2}\right) &= (1-ie^{\pi b^{\pm1}x})e_\R\left(x+\frac{ib^{\pm1}}{2}\right).
\end{align*}
Functions $e_\NS(x)$ and $e_\R(x)$ satisfying these properties can be constructed as 
\begin{align}
 e_\R(x) &= e_\ub\left(\frac{x+i(b-b^{-1})\slash2}{2}\right)e_\ub\left(\frac{x-i(b-b^{-1})\slash2}{2}\right),\\
 e_\NS(x) &= e_\ub\left(\frac{x+c_\ub}{2}\right)e_\ub\left(\frac{x-c_\ub}{2}\right),
\end{align}
where $e_\ub(x)$ is Faddeev's quantum dilogarithm function defined by the following integral representation
\begin{equation}
 e_\ub(x) = \exp \left[\int_{\mathbb{R}+i0} \ud w \frac{e^{-2ixw}}{4 \sinh(wb)\sinh(w\slash b)}\right],
\end{equation}
Some details on the verification of the quantised coordinate transformations (\ref{flipoperator})
are given in appendix \ref{appendix3}. 

As a useful tool for describing the definition
of the remaining operators 
$\mathsf{T}^{(i)}_{12}$, $i=2,\dots,8$, we will introduce an operator 
$\mathsf{M}_v: \mathcal{H}_v\to \mathcal{H}_v$ representing the change of orientations $\mu_v$ in an undotted triangle 
shown in the figure \ref{operatorMM}. The operator $\mathsf{M}_v$ is associated by our conventions concerning 
tensor products introduced above to the operator $\mathsf{M}$ on $\mathbb{C}^{1|1}$ which can be represented
by the matrix
\begin{equation}
 \mathsf{M}= \bigg( \begin{array}{cc} 1 & 0 \\ 0 & -1 \end{array} \bigg).
\end{equation}
The operator $\mathsf{M}_v$ squares to identity $\mathsf{M}^2_v = {\rm id}_v$ and acts on the odd invariant as
 \begin{equation}\label{U}
  \mathsf{M}_v^{-1}\cdot \xi_v^{}\cdot \mathsf{M}^{}_v = -\xi_v^{}.
 \end{equation} 
One should note that the operation $\mu_v$ relates Kasteleyn orientations describing inequivalent spin structures, 
in general. 
 
\begin{figure}[h]
\centering
\includegraphics[width=0.6\textwidth]{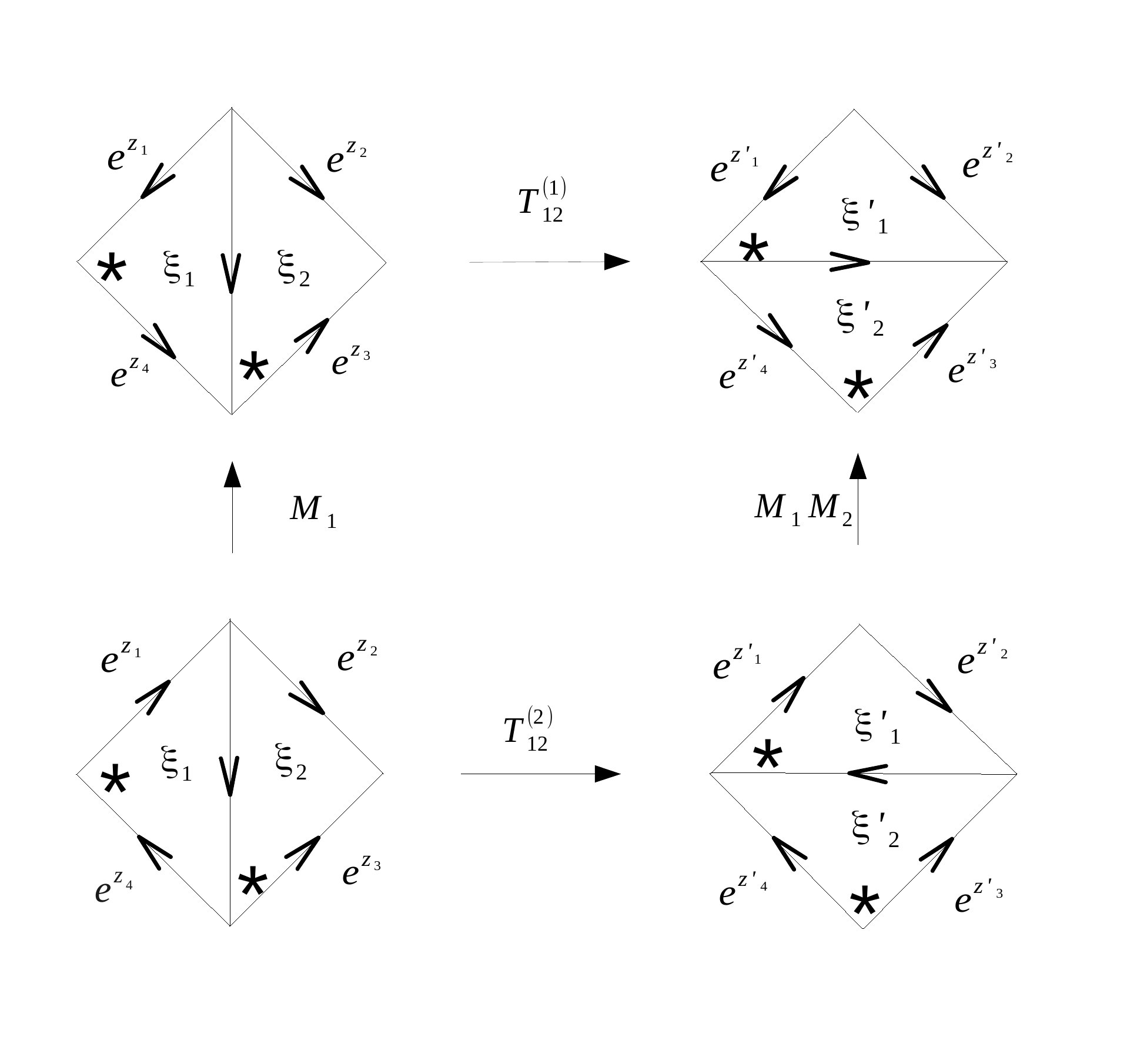}
\caption{By using operators $\mathsf{M}$ we can find the map between the second superflip and the first one.}
\label{exampleT1T2}
\end{figure}

It is easy to see that the flips $\omega^{(i)}_{12}$, $i=2,\dots,8$ can be represented as 
compositions of the flip $\omega^{(1)}_{12}$ with 
operations $\mu_v$. We will define the corresponding operators
$\mathsf{T}^{(i)}_{12}$, $i=2,\dots,8$ 
by taking the corresponding product of the operators $\mathsf{M}_v$ with the operator $\mathsf{T}^{(1)}_{12}$. 
To give an example, let us note that the flip  $\omega^{(2)}$ can be represented by the sequence of 
operations shown in figure \ref{exampleT1T2}. This leads us to define the operator $\mathsf{T}^{(2)}_{12}$
as
\begin{equation}
\mathsf{T}_{12}^{(2)}=\mathsf{M}_1\mathsf{M}_2\mathsf{T}_{12}^{(1)}\mathsf{M}_1 .
\end{equation}
All other operators $\mathsf{T}^{(i)}_{12}$, $i=3,\dots,8$ associated to the flips 
$\omega^{(i)}$, $i=3,\dots,8$ can be defined in this way.

The operations considered up to now were associated to triangles that do not have corners marked with 
dots. As noted above, one may always locally reduce to this case by using the push-out operation.
The push-out $\beta$ will be 
represented by an operator $\mathsf{B}_{uv}:\mathcal{H}_u\otimes\mathcal{H}_v\to\mathcal{H}_u\otimes\mathcal{H}_v$ defined as follows
 \begin{equation}
   \mathsf{B}_{uv} = {\rm id}_u\, \mathsf{M}_v.
 \end{equation}
With the help of the operator $\mathsf{B}_{uv}$ one may now define all operators 
associated with the flips relating dotted triangles. 

We furthermore need to define operators
$\Pi^{(i)}_{(12)}$, $i=1,\dots,8$ representing the exchange $(uv)$ of labels assigned to two adjacent triangles 
when the Kastelyn orientation is the one of the initial configurations of the flips $\omega_{12}^{(i)}$ depicted in 
Figure \ref{superflips1}. By using the operators $\mathsf{M}_v$ one may reduce the definition to the case 
$i=1$ in a way closely analogous to the definition of the $\mathsf{T}_{12}^{(i)}$, $i=2,\dots,8$ 
in terms  of $\mathsf{T}_{12}^{(1)}$. In order to define the 
operator $\Pi^{(1)}_{(12)}$ let us represent $\mathcal{H}_1\otimes\mathcal{H}_2$ as 
$L^2(\mathbb{R}^2)\otimes \mathbb{C}^{1|1}\otimes  \mathbb{C}^{1|1}$, and let
\begin{equation}\label{permdef}
 \Pi^{(1)}_{(12)} =  ( \mathsf{P}_{\rm b} \otimes\mathbb{I}_2\otimes\mathbb{I}_2)({\rm id}\otimes 
 \mathsf{P}_{\rm f})\,,\quad\text{where}\quad
 \mathsf{P}_{\rm f}=
 (\mathbb{I}_2\otimes \mathsf{M}) (\mathbb{I}_2\otimes\mathbb{I}_2 + \kappa\otimes\kappa ),
\end{equation}
with respect to this factorisation, where
$\mathsf{P}_{\rm b}$ acts on  functions of two variables as
$ \mathsf{P}_{\rm b} f(x_1,x_2) = f(x_2,x_1)$. One may note that  
$\mathsf{P}_{\rm f}$ is not the standard permutation operator on 
$\mathbb{C}^{1|1}\otimes\mathbb{C}^{1|1}$ satisfying $ \mathsf{P}_{\rm f} (\eta_1\otimes\eta_2)\mathsf{P}_{\rm f}
=\eta_2\otimes \eta_1$
for arbitrary $\eta_1,\eta_2\in{\rm End}(\mathbb{C}^{1|1})$. However, 
the operator $\mathsf{P}_{\rm f}$ squares to the identity 
and satisfies $ \mathsf{P}_{\rm f} (\xi\otimes\mathbb{I}_2)\mathsf{P}_{\rm f}=\mathbb{I}_2\otimes \xi$
and $ \mathsf{P}_{\rm f} ( \mathbb{I}_2\otimes \xi)\mathsf{P}_{\rm f}=\xi\otimes\mathbb{I}_2$. This means that
the operator $\mathsf{P}_{\rm f}$ correctly represents the permutation on the the sub-algebra
of ${\rm End}(\mathbb{C}^{1|1}\otimes\mathbb{C}^{1|1})$ generated by $\mathbb{I}_2\otimes \xi$ and 
$\xi\otimes \mathbb{I}_2$. This is the algebra of operators on  $\mathbb{C}^{1|1}\otimes\mathbb{C}^{1|1}$
relevant for
the quantisation of the super-\TM theory. The reason for adopting a non-standard representation of the 
permutation on this sub-algebra will become clear when we discuss the relations of the 
super Ptolemy groupoid.

We finally need to define an operator $\mathsf{A}_v$ representing the 
move rotating the distinguished vertex of a dotted triangle as shown in figure \ref{classicalmap2}. 
The operator $\mathsf{A}_v: \mathcal{H}_v\to\mathcal{H}_v$  will be defined as
\begin{equation}
 \mathsf{A}_v = e^{i\pi/3} e^{-i3\pi \mathsf{q}_v^2\slash2}e^{-i\pi(\mathsf{p}_v+\mathsf{q}_v)^2\slash2} \mathbb{I}_2.
\end{equation}

Let us finally note that 
the  flip operators $\mathsf{T}^{(i)}_{12}$ have an interesting interpretation within the 
representation theory of the Heisenberg double of the quantum super plane, 
which will be elaborated in the forthcoming paper \cite{APT_heisenbergdouble}. 
The flip operator $\mathsf{T}^{(1)}_{12}$ is found to coincide with the canonical element of the 
Heisenberg double of the quantum super plane (which is a Borel half of $U_q(osp(1|2))$), 
evaluated in certain infinite-dimensional representations on $L^2(\mathbb{R})\otimes \mathbb{C}^{1|1}$. 

\subsection{Quantum super Ptolemy groupoid}

We are now going to describe essential steps in the verification that the operators defined
previously generate a representation of the super Ptolemy groupoid.

\begin{figure}
\centering
\includegraphics[width=0.6\textwidth]{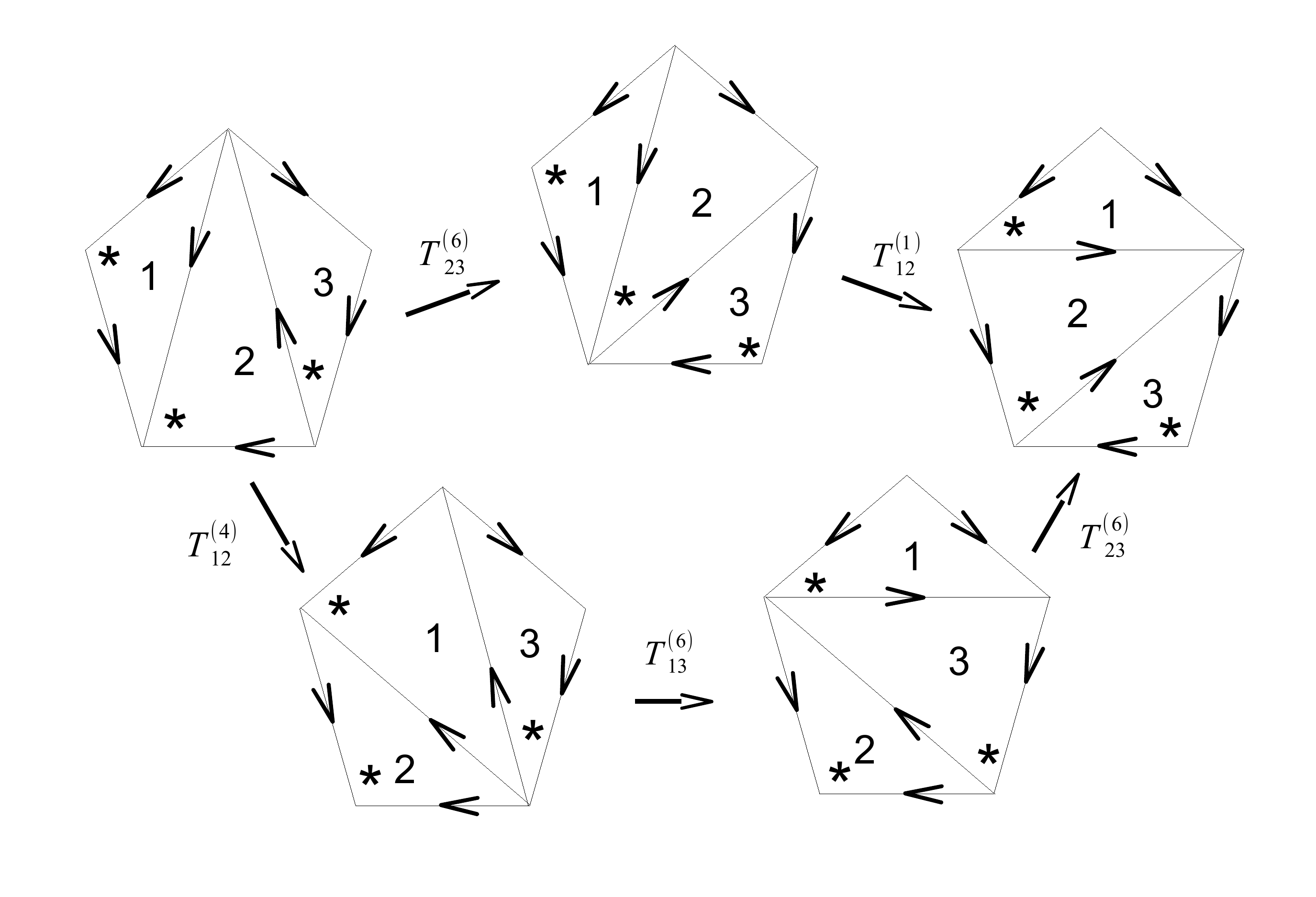}
\caption{One of the pentagon equations.}
\label{superpentagonquantumexample}
\end{figure}

Of particular interest are the generalisations of the pentagon relation. Using the push-out operation 
one can always reduce to relations involving only undotted triangles. As noted previously, one needs to 
check the following set of relations,
\begin{align}\label{superpentagonq}
\begin{split}
\mathsf{T}^{(1)}_{12} \mathsf{T}^{(1)}_{23}= \mathsf{T}^{(1)}_{23} \mathsf{T}^{(1)}_{13} \mathsf{T}^{(1)}_{12}, &\qquad \mathsf{T}^{(6)}_{12} \mathsf{T}^{(2)}_{23} = \mathsf{T}^{(2)}_{23} \mathsf{T}^{(1)}_{13} \mathsf{T}^{(6)}_{12},\\
\mathsf{T}^{(5)}_{12} \mathsf{T}^{(8)}_{23}= \mathsf{T}^{(8)}_{23} \mathsf{T}^{(5)}_{13} \mathsf{T}^{(5)}_{12}, &\qquad  \mathsf{T}^{(6)}_{12} \mathsf{T}^{(7)}_{23} = \mathsf{T}^{(7)}_{23} \mathsf{T}^{(6)}_{13} \mathsf{T}^{(5)}_{12},\\
\mathsf{T}^{(2)}_{12} \mathsf{T}^{(1)}_{23}= \mathsf{T}^{(1)}_{23} \mathsf{T}^{(2)}_{13} \mathsf{T}^{(2)}_{12}, &\qquad \mathsf{T}^{(8)}_{12} \mathsf{T}^{(8)}_{23} = \mathsf{T}^{(1)}_{23} \mathsf{T}^{(8)}_{13} \mathsf{T}^{(8)}_{12},\\
\mathsf{T}^{(4)}_{12} \mathsf{T}^{(5)}_{23}= \mathsf{T}^{(5)}_{23} \mathsf{T}^{(5)}_{13} \mathsf{T}^{(4)}_{12},&\qquad \mathsf{T}^{(5)}_{12} \mathsf{T}^{(3)}_{23} = \mathsf{T}^{(3)}_{23} \mathsf{T}^{(4)}_{13} \mathsf{T}^{(6)}_{12},\\
\mathsf{T}^{(3)}_{12} \mathsf{T}^{(4)}_{23}= \mathsf{T}^{(7)}_{23} \mathsf{T}^{(3)}_{13} \mathsf{T}^{(2)}_{12},&\qquad \mathsf{T}^{(7)}_{12} \mathsf{T}^{(7)}_{23} = \mathsf{T}^{(4)}_{23} \mathsf{T}^{(7)}_{13} \mathsf{T}^{(8)}_{12},\\
\mathsf{T}^{(1)}_{12} \mathsf{T}^{(6)}_{23}= \mathsf{T}^{(6)}_{23} \mathsf{T}^{(6)}_{13} \mathsf{T}^{(4)}_{12},
&\qquad \mathsf{T}^{(7)}_{12} \mathsf{T}^{(2)}_{23} = \mathsf{T}^{(5)}_{23} \mathsf{T}^{(2)}_{13} \mathsf{T}^{(7)}_{12},\\
\mathsf{T}^{(5)}_{12} \mathsf{T}^{(6)}_{23}= \mathsf{T}^{(3)}_{23} \mathsf{T}^{(7)}_{13} \mathsf{T}^{(6)}_{12},&\qquad \mathsf{T}^{(3)}_{12} \mathsf{T}^{(5)}_{23} = \mathsf{T}^{(2)}_{23} \mathsf{T}^{(8)}_{13} \mathsf{T}^{(3)}_{12},\\
\mathsf{T}^{(1)}_{12} \mathsf{T}^{(3)}_{23}= \mathsf{T}^{(6)}_{23} \mathsf{T}^{(3)}_{13} \mathsf{T}^{(7)}_{12},&\qquad \mathsf{T}^{(4)}_{12} \mathsf{T}^{(4)}_{23} = \mathsf{T}^{(4)}_{23} \mathsf{T}^{(4)}_{13} \mathsf{T}^{(1)}_{12}.
\end{split}
\end{align}
One may first observe that all of these relations follow 
from the pentagon equation that involves only $\mathsf{T}^{(1)}$.
As an example let us consider the pentagon equation represented by
Figure \ref{superpentagonquantumexample}, corresponding to the equation
 \begin{equation*}
  \mathsf{T}^{(6)}_{12} \mathsf{T}^{(2)}_{23} = \mathsf{T}^{(2)}_{23} \mathsf{T}^{(1)}_{13} \mathsf{T}^{(6)}_{12}.
 \end{equation*}
 Using the relations between $\mathsf{T}^{(1)}$ and other flips, we can rewrite it 
\begin{equation*}
  (\mathsf{M}_2 \mathsf{T}^{(1)}_{12} \mathsf{M}_1\mathsf{M}_2) (\mathsf{M}_2\mathsf{M}_3 \mathsf{T}^{(1)}_{23} \mathsf{M}_2) = (\mathsf{M}_2\mathsf{M}_3 \mathsf{T}^{(1)}_{23} \mathsf{M}_2) \mathsf{T}^{(1)}_{13} (\mathsf{M}_2 \mathsf{T}^{(1)}_{12} \mathsf{M}_1\mathsf{M}_2),
\end{equation*}
which is just a pentagon for $\mathsf{T}^{(1)}$, given the fact that 
$\mathsf{M}_1\mathsf{M}_2 \mathsf{T}^{(i)}_{12} \mathsf{M}_1\mathsf{M}_2 = \mathsf{T}^{(i)}_{12}$ for all $i$.

In order to verify 
the pentagon equation for $\mathsf{T}^{(1)}$ 
one may note that by straightforward calculations one may reduce the validity of this relations to  
the following identities 
\begin{subequations} \label{oppenta} \begin{align}
    f_+(\mathsf{p}) f_+(\mathsf{x}) &=  f_+(\mathsf{x})f_+(\mathsf{x}+\mathsf{p})f_+(\mathsf{p}) -i f_-(\mathsf{x})f_-(\mathsf{x}+\mathsf{p})f_-(\mathsf{p}), \\
   f_+(\mathsf{p}) f_-(\mathsf{x}) &=  -i f_+(\mathsf{x})f_-(\mathsf{x}+\mathsf{p})f_-(\mathsf{p}) + f_-(\mathsf{x})f_+(\mathsf{x}+\mathsf{p})f_+(\mathsf{p}), \\
   f_-(\mathsf{p}) f_+(\mathsf{x}) &=  f_+(\mathsf{x}) f_+(\mathsf{x}+\mathsf{p}) f_-(\mathsf{p}) -i f_-(\mathsf{x})f_-(\mathsf{x}+\mathsf{p})f_+(\mathsf{p}) , \\
   f_-(\mathsf{p}) f_-(\mathsf{x}) &=  i f_+(\mathsf{x}) f_-(\mathsf{x}+\mathsf{p}) f_+(\mathsf{p}) - f_-(\mathsf{x})f_+(\mathsf{x}+\mathsf{p})f_-(\mathsf{p}),
 \end{align}\end{subequations}
which are valid if the operators $\mathsf{x}$ and $\mathsf{p}$ that satisfy
\begin{equation*}
 [\mathsf{p},\mathsf{x}] = \frac{1}{i\pi}.
\end{equation*}
The relations (\ref{oppenta}) follow from integral identities satisfied by the 
special  functions
$e_\NS(x)$ and $e_\R(x)$ that were derived in \cite{Hadasz:2007wi}, see Appendix \ref{pentproof}
for details.

\begin{figure}[h]
\centering
\includegraphics[width=0.9\textwidth]{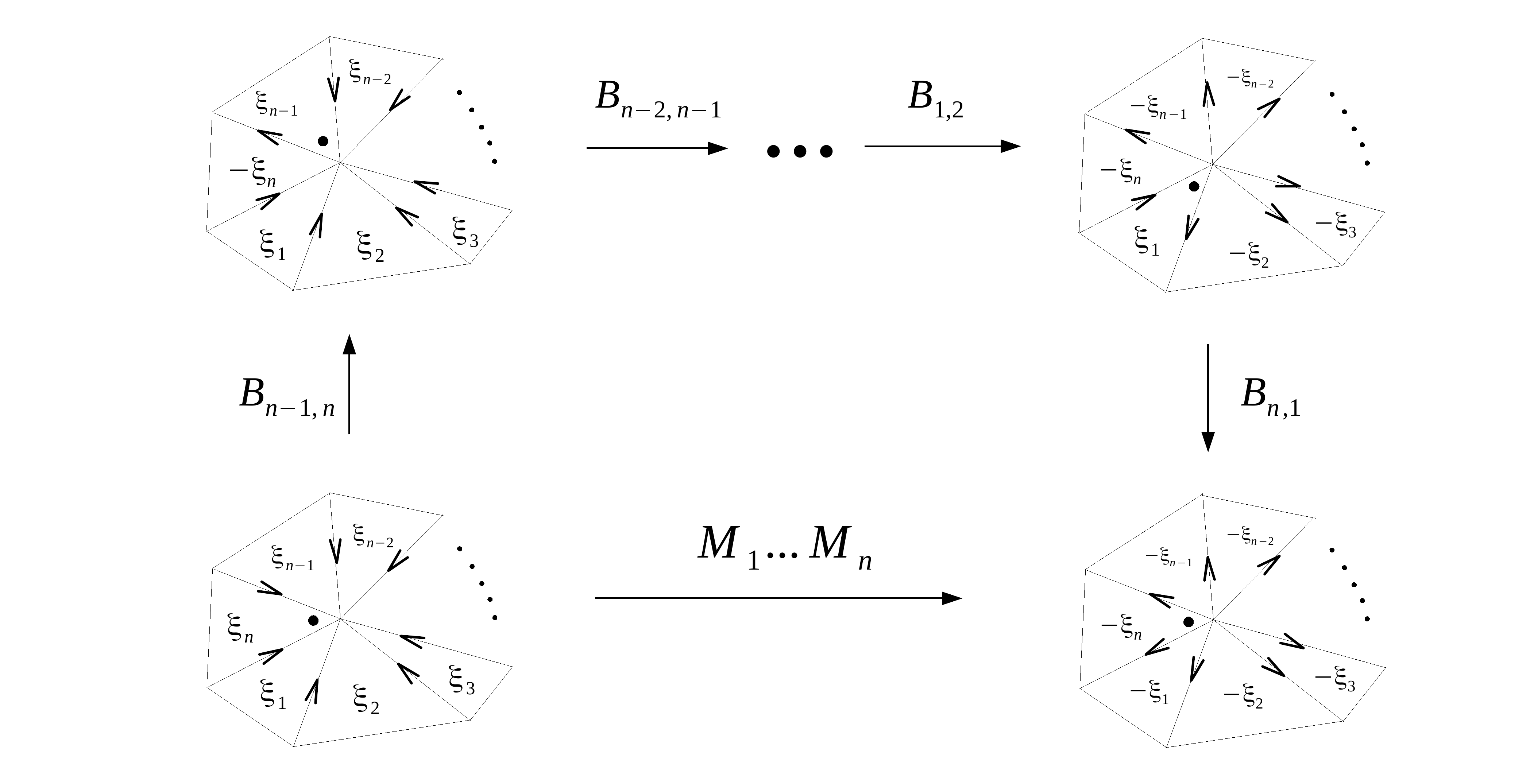}
\caption{Relation for push-out.}
\label{pushoutoperator}
\end{figure}

The quantum Ptolemy groupoid is defined by relations besides the super pentagon. There is an equations satisfied by a push-out
\begin{equation}
 \mathsf{B}_{n,1} \mathsf{B}_{1,2} \ldots \mathsf{B}_{n-1,n} = \mathsf{M}_1\mathsf{M}_2\cdots \mathsf{M}_n ,
\end{equation}
for all $n\geq2$, which comes from the figure \ref{pushoutoperator}, where we consider a collection of hexagons meeting in the same vertex (a collection of vertices in $S^{1|1}$ that project to the same point in $\mathbb{P}^{1|1}$). Then, we can move the dot around this vertex until we arrive at the same hexagon, and then relate this hexagonalisation to the initial one by reversing the orientation on the edges.

Moreover, we want to derive the relations between flips and push-outs. In order to find a minimal number of equations which one could use as a defining relations for super Ptolemy groupoid, we consider the collections of hexagons, with one dot (by sequences of push-outs one can reduce the cases of multiple dots to the case of one dot). Therefore we have
 \begin{equation}\label{pushouts2flips1}
 (\mathsf{T}^{(i)}_{23})^{-1} \mathsf{B}_{43}^{} \mathsf{B}_{32}^{}
\mathsf{B}_{21}^{}= \mathsf{B}_{42}^{} \mathsf{B}_{21}^{}(\mathsf{T}^{(j)}_{23})^{-1},
\end{equation}
where the pairs $(i,j) = (5,8),(8,5),(6,7),(7,6),(1,2),(2,1),(3,4),(4,3)$, 
\begin{equation}
 \mathsf{T}^{(i)}_{23} \mathsf{B}_{43}^{} \mathsf{B}_{32}^{} \mathsf{B}_{21}^{} = \mathsf{B}_{43}^{} \mathsf{B}_{31}^{} \mathsf{T}^{(j)}_{23},
\end{equation}
where the pairs $(i,j) = (5,4),(4,5),(1,6),(6,1),(7,2),(2,7),(3,8),(8,3)$, 
\begin{align}
&\mathsf{T}_{34}^{(i)} \mathsf{B}_{23}^{} (\mathsf{T}_{12}^{(j)})^{-1} (\mathsf{B}_{23}^{})^{-1} = \mathsf{B}_{24}^{} (\mathsf{T}_{12}^{(j)})^{-1} (\mathsf{B}_{24}^{})^{-1} \mathsf{T}_{34}^{(i)}, \\
&(\mathsf{T}_{34}^{(i)})^{-1} \mathsf{B}_{13}^{} \mathsf{T}_{12}^{(j)} (\mathsf{B}_{23}^{})^{-1} = \mathsf{B}_{13}^{} \mathsf{T}_{12}^{(j)} (\mathsf{B}_{23}^{})^{-1} (\mathsf{T}_{34}^{(i)})^{-1},\\
&(\mathsf{T}_{34}^{(i)})^{-1} \mathsf{B}_{23}^{} (\mathsf{T}_{12}^{(j)})^{-1} (\mathsf{B}_{23}^{})^{-1} = \mathsf{B}_{23}^{} (\mathsf{T}_{12}^{(j)})^{-1} (\mathsf{B}_{23}^{})^{-1} (\mathsf{T}_{34}^{(i)})^{-1},\\
&\mathsf{T}_{34}^{(i)} \mathsf{B}_{13}^{} \mathsf{T}_{12}^{(j)} (\mathsf{B}_{23}^{})^{-1} = \mathsf{B}_{14}^{} \mathsf{T}_{12}^{(j)} (\mathsf{B}_{24}^{})^{-1} \mathsf{T}_{34}^{(i)},\label{pushouts4flips4}
\end{align}
where the $i,j,k,l,m=1,\ldots,8$ depends on the Kasteleyn orientation of the graph from which the relation has been derived. Examples of these 
relations are represented diagrammatically in figures \ref{dot1example} and \ref{dot2example}, with decorated vertices assigned appropriately. 

We finally need to discuss the relations of the super Ptolemy groupoid  involving the operator  $\mathsf{A}$. We find 
that the following relations are satisfied
\begin{align}\label{rotationsandflips1}
 & \mathsf{A}^3_1 = {\rm id}_1, \\
 & \mathsf{A}_2 \mathsf{T}^{(i)}_{12} \mathsf{A}_1 = \mathsf{A}_1 \mathsf{T}^{(i)}_{21} \mathsf{A}_2, \\
 &  \mathsf{T}^{(j)}_{21} \mathsf{A}_1 \mathsf{T}^{(k)}_{12} = \zeta_s\,\mathsf{A}^{}_2 \mathsf{A}^{}_1\Pi^{(k)}_{(12)} 
 ,\label{rotationsandflips3}
\end{align}
where $i=1,\ldots,8$, the pairs $(j,k)=(4,1),(7,2),(2,3),(5,4),(8,5),(3,6),(6,7),(1,8)$, and 
$\zeta_s=e^{\frac{\pi i}{4}} e^{-i\pi(1+ c_\ub^2)\slash6}$.
Details on the proof of (\ref{rotationsandflips3}) 
can be found in Appendix \ref{appendixD}. It is the operator $\Pi^{(1)}_{(12)} $
 defined in \eqref{permdef} which 
appears in  (\ref{rotationsandflips3}) for $i=1$, explaining why we adopted this definition for $\Pi^{(1)}_{(12)}$.



\section{Conclusions and outlook}
In this work we constructed a quantisation of the \TM theory of super
Riemann surfaces.  The independence of the resulting quantum theory with respect
to changes of triangulations 
was demonstrated by constructing a unitary projective representation of the 
super Ptolemy groupoid. 

There is a number of issues which would be interesting to investigate. It is
known that ordinary \TM theory is closely related to non-supersymmetric
Liouville theory \cite{T14a}. In particular, the spaces of Liouville conformal
blocks and the spaces of states of \TM theory of Riemann surfaces can
be identified as predicted in \cite{Verlinde} and they carry unitarily equivalent
representations of the mapping class group. In the case of $N=1$ supersymmetric Liouville theory, 
the mapping class group representation for genus $0$ represented using the 
fusion and braiding matrices, has been investigated
\cite{Chorazkiewicz:2008es,Chorazkiewicz:2011zd}. It would be
interesting to study more closely the mapping class group
representation defined  by the representation of the 
super Ptolemy groupoid constructed in our paper, and to relate it to
$N=1$ supersymmetric Liouville theory. 

Moreover, ordinary \TM theory is the connected component of the space
of $SL(2,\mathbb{R})$-valued flat connections on a Riemann surface
$\Sigma$, and therefore closely related to
$SL(2,\mathbb{R})$-Chern-Simons theory on $\Sigma\times\mathbb{R}$.
It should be interesting
to investigate the connections between the quantum super \TM
theory described here and the quantum $OSp(1|2)$-Chern-Simons theory.\\

\smallskip

\noindent
{\bf Acknowledgements:} We would like to thank Fabien Bouschbacher, Vladimir Fock, Ivan Ip, Sebastian Novak and Volker Schomerus for useful discussions.

The work of Nezhla Aghaei and Michal Pawelkiewicz was supported by the German Science Foundation (DFG) within the Research Training Group 1670 ``Mathematics Inspired by String Theory and QFT''. Support from the DFG in the framework of the SFB 676 \emph{Particles Strings, and the Early Universe} is gratefully acknowledged.

\appendix

\section{Special functions}\label{appendix1}

\subsection{Non-compact quantum dilogarithm}

\label{def_specfunc}

The basic special function that appears in the context
of the quantisation of the \TM space is Barnes' double Gamma function. For
$\mathfrak{Re}x>0$ it admits an integral representation
\begin{eqnarray*}
 \log \Gamma_\ub(x) = \int_0^\infty \frac{\ud t}{t}
 \left[ \frac{e^{-x t} - e^{-\frac{Q}{2} t}}{(1 - e^{-t \ub})
 (1 - e^{-\frac{t}{\ub}})} - \frac{\left( \frac{Q}{2}-x\right)^2}{2 e^t}
 - \frac{\frac{Q}{2} - x}{t} \right] ,
\end{eqnarray*}
where $Q = \ub + \frac{1}{\ub}$. One can analytically continue $\Gamma_\ub$
to a meromorphic function defined on the entire complex plane
$\mathbb{C}$. The most important property of $\Gamma_\ub$ is its
behavior with respect to shifts by $\ub^\pm$,
\begin{equation}
 \Gamma_\ub(x+\ub) = \frac{\sqrt{2\pi} \ub^{\ub x-\frac{1}{2}}}{\Gamma_\ub(bx)}
 \Gamma_\ub(x)\quad , \quad
 \Gamma_\ub(x+\ub^{-1}) = \frac{\sqrt{2\pi} \ub^{-\frac{\ub}{x}+\frac{1}{2}}}
 {\Gamma_\ub(\frac{x}{\ub})} \Gamma_\ub(x)\ .
\end{equation}
These shift equation allows us to calculate residues of the poles of
$\Gamma_\ub$. When $x\to0$, for instance, one finds
\begin{equation}\label{residue}
 \Gamma_\ub(x)= \frac{\Gamma_\ub(Q)}{2\pi x} + O(1).
\end{equation}
From Barnes' double Gamma function we can build two other important
special functions,
\begin{align}
 S_\ub(x) &= \frac{\Gamma_\ub(x)}{\Gamma_\ub(Q-x)}, \\[2mm]
 G_\ub(x) &= e^{-\frac{i\pi}{2} x(Q-x)} S_\ub(x).\label{definition_G}
\end{align}
We shall often refer to the function $S_\ub$ as double sine function. The $S_\ub$ function is meromorphic with poles and zeros in
\begin{align*}
  S_\ub(x) = 0 &\Leftrightarrow x = Q + n b + m b^{-1}, \quad n,m \in \mathbb{Z}_{\geq0}\ , \\[2mm]
  S_\ub(x)^{-1} = 0 &\Leftrightarrow x = -n b -m b^{-1}, \quad n,m \in \mathbb{Z}_{\geq0}\ .
\end{align*}
From its definition and the shift property of Barnes' double Gamma
function it is easy to derive the following shift and reflection
properties of $G_\ub$,
\begin{align}
 &G_\ub(x+\ub) = (1-e^{2 \pi i \ub x}) G_\ub(x)\ ,\label{shift} \\[2mm]
 &G_\ub(x) G_\ub(Q-x) = e^{\pi i x(x-Q)}\ .\label{reflection}
\end{align}
The Faddeev's quantum dilogarithm function is defined by the following integral representation
\begin{equation}
 e_\ub(x) = \exp \left[\int_{\mathbb{R}+i0} \ud w \frac{e^{-2ixw}}{4 \sinh(wb)\sinh(w\slash b)}\right],
\end{equation}
and it is related to the double sine function in a way as follows
\begin{equation}
 e_\ub(x) =\ A G_\ub^{-1}(-ix + \frac{Q}{2}),
\end{equation}
where
\begin{equation} \label{eq:A}
 A \, = \, e^{-i\pi(1-4 c_\ub^2)\slash 12}\quad , \quad  c_\ub = i Q/2\ .
\end{equation}
The shift and reflection relations that it satisfies are as follows
\begin{align*}
 e_\ub\left(x-\frac{i b^{\pm1}}{2}\right) &= (1+e^{2\pi b^{\pm1}x})e_\ub\left(x+\frac{ib^{\pm1}}{2}\right), \\
 e_\ub(x)e_\ub(-x) &= e^{-i\pi(1+2c_b^2)\slash 6} e^{i\pi x^2}.
\end{align*}
The asymptotic behaviour of the function $e_\ub$ along the real axis 
\begin{equation} \label{asymptotic}
 e_\ub(z) = \left\{ \begin{array}{ll}
 1 &, x\to -\infty\\
  e^{-i\pi(1+2c_b^2)\slash 6} e^{i\pi x^2} & ,x\to +\infty\\
\end{array} \right.
\end{equation}
Also, we know that for self-adjoint operators 
$\mathsf{P},\mathsf{X}$ such that $[\mathsf{P},\mathsf{X}]=\frac{1}{2\pi i}$  we have the following 
variant of the pentagon relation 
 \begin{equation}
  e_\ub(\mathsf{P}) e_\ub(\mathsf{X}) = e_\ub(\mathsf{X}) e_\ub(\mathsf{X}+\mathsf{P}) e_\ub(\mathsf{P}) .
 \end{equation}
The pentagon equation is equivalent to the following 
analog of the  Ramanujan summation formula \cite{PT2,FKV,Vol}
\begin{equation} \label{Ramanujan}
 \int_{-i\infty}^{i\infty} \frac{\ud\tau}{i} e^{2 \pi i \tau \beta} \frac{G_\ub(\tau + \alpha)}{G_\ub(\tau + Q)} = \frac{G_\ub(\alpha) G_\ub(\beta)}{G_\ub(\alpha+\beta)}.
\end{equation}
It may also be considered  as a quantisation of the Rogers five-term identity satisfied by dilogarithms. 
  
  \subsection{Supersymmetric non-compact quantum dilogarithm}

When discussing the supersymmetric \TM theory we need the following additional special functions
\begin{align*}
 \Gamma_1(x) & = \Gamma_\NS(x) = \Gamma_\ub\left(\frac{x}{2}\right) \Gamma_\ub
 \left(\frac{x+Q}{2}\right),\\[2mm]
 \Gamma_0(x) & = \Gamma_\R(x) = \Gamma_\ub\left(\frac{x+\ub}{2}\right) \Gamma_\ub
 \left(\frac{x+\ub^{-1}}{2}\right).
\end{align*}
Furthermore, let us define
\begin{equation}
\begin{array}{rlrl}
 S_1(x)&\!= S_\NS(x) = \frac{\Gamma_\NS(x)}{\Gamma_\NS(Q-x)} , \quad &
G_1(x)&\!= G_\NS(x) = \zeta_0 e^{-\frac{i\pi}{4} x(Q-x)} S_\NS(x), \\[2mm]
  S_0(x)&\!= S_\R(x) = \frac{\Gamma_\R(x)}{\Gamma_\R(Q-x)}, \quad &
 G_0(x)&\!= G_\R(x) = e^{-\frac{i\pi}{4}} \zeta_0 e^{-\frac{i\pi}{4}
 x(Q-x)} S_\R(x),
\end{array}
\end{equation}
where $\zeta_0 = \exp(-i\pi Q^2/8)$. As for $S_\ub$, the functions $S_0(x)$ and $S_1(x)$ are meromorphic with poles and zeros in
\begin{align*}\
  S_0(x) = 0 &\Leftrightarrow x = Q + n b + m b^{-1}, \quad n,m \in \mathbb{Z}_{\geq0}, m+n\in2\mathbb{Z}+1, \\
  S_1(x) = 0 &\Leftrightarrow x = Q + n b + m b^{-1}, \quad n,m \in \mathbb{Z}_{\geq0}, m+n\in2\mathbb{Z},\\
  S_0(x)^{-1} = 0 &\Leftrightarrow x = -n b -m b^{-1}, \quad n,m \in \mathbb{Z}_{\geq0}, m+n\in2\mathbb{Z}+1, \\
  S_1(x)^{-1} = 0 &\Leftrightarrow x = -n b -m b^{-1}, \quad n,m \in \mathbb{Z}_{\geq0}, m+n\in2\mathbb{Z}.
\end{align*}
As in the previous subsection, we want to state the shift and reflection properties of the functions $G_1$
and $G_0$,
\begin{align}
  G_\nu(x+b^{\pm1}) &= (1-(-1)^\nu e^{ \pi i \ub^{\pm1} x}) G_{\nu+1}(x),
  \label{shift_super}\\[2mm]
  \label{reflection_super}
  G_\nu(x) G_\nu(Q-x) &= e^{\frac{i\pi}{2}(\nu-1)} \zeta_0^2
  e^{\frac{\pi i}{2} x(x-Q)}\ .
\end{align}
We define the supersymmetric analogues of Faddeev's quantum dilogarithm function as
\begin{align}
 e_\R(x) &= e_\ub\left(\frac{x+i(b-b^{-1})\slash2}{2}\right)e_\ub\left(\frac{x-i(b-b^{-1})\slash2}{2}\right),\\
 e_\NS(x) &= e_\ub\left(\frac{x+c_\ub}{2}\right)e_\ub\left(\frac{x-c_\ub}{2}\right),
\end{align}
and relate them to the double sine function in a way as follows
\begin{equation}
 e_\nu(x) =\ A^2 G_\nu^{-1}(-ix + \frac{Q}{2}),
\end{equation}
with a constant $A$ as defined in eq.\ \eqref{eq:A}.  
The shift and reflection relations that it satisfies are as follows
\begin{align*}
 e_\R\left(x-\frac{i b^{\pm1}}{2}\right) &= (1+ie^{\pi b^{\pm1}x})e_\NS\left(x+\frac{ib^{\pm1}}{2}\right), \\
 e_\NS\left(x-\frac{i b^{\pm1}}{2}\right) &= (1-ie^{\pi b^{\pm1}x})e_\R\left(x+\frac{ib^{\pm1}}{2}\right), \\
 e_\NS(x)e_\NS(-x) &= e^{i\pi c_\ub^2\slash2} e^{-i\pi(1+2c_b^2)\slash 3} e^{i\pi x^2\slash 2}, \\
 e_\R(x)e_\R(-x) &= e^{i\pi\slash 2} e^{i\pi c_\ub^2\slash2} e^{-i\pi(1+2c_b^2)\slash 3} e^{i\pi x^2\slash 2}.
\end{align*}
Asymptotically, the functions $e_1$ and $e_0$ behave as
\begin{align}
\label{asymptotic_superNS}
 e_\NS(z) &= \left\{ \begin{array}{ll}
 1 &, x\to -\infty\\
 e^{i\pi c_\ub^2\slash2} e^{-i\pi(1+2c_b^2)\slash 3} e^{i\pi x^2\slash 2} & ,x\to +\infty\\
\end{array} \right.\\
 e_\R(z) &= \left\{ \begin{array}{ll}
 1 &, x\to -\infty\\
 e^{i\pi\slash 2} e^{i\pi c_\ub^2\slash2} e^{-i\pi(1+2c_b^2)\slash 3} e^{i\pi x^2\slash 2} & ,x\to +\infty\\
\end{array} \right.
\end{align}
Also, we know that for self-adjoint operators $\mathsf{P},\mathsf{X}$ such that $[\mathsf{P},\mathsf{X}]=\frac{1}{ \pi i}$ they satisfy four pentagon relations
\begin{subequations}\label{superpentagon}
 \begin{align}
    f_+(\mathsf{P}) f_+(\mathsf{X}) &=  f_+(\mathsf{X})f_+(\mathsf{X}+\mathsf{P})f_+(\mathsf{P}) -i f_-(\mathsf{X})f_-(\mathsf{X}+\mathsf{P})f_-(\mathsf{P}), \\
     f_+(\mathsf{P}) f_-(\mathsf{X}) &=  -i f_+(\mathsf{X})f_-(\mathsf{X}+\mathsf{P})f_-(\mathsf{P}) + f_-(\mathsf{X})f_+(\mathsf{X}+\mathsf{P})f_+(\mathsf{P}), \\
     f_-(\mathsf{P}) f_+(\mathsf{X}) &=  f_+(\mathsf{X}) f_+(\mathsf{X}+\mathsf{P}) f_-(\mathsf{P}) -i f_-(\mathsf{X})f_-(\mathsf{X}+\mathsf{P})f_+(\mathsf{P}) , \\
     f_-(\mathsf{P}) f_-(\mathsf{X}) &=  i f_+(\mathsf{X}) f_-(\mathsf{X}+\mathsf{P}) f_+(\mathsf{P}) - f_-(\mathsf{X})f_+(\mathsf{X}+\mathsf{P})f_-(\mathsf{P}),
 \end{align}
 \end{subequations}
where $f_\pm(x) = e_\R(x) \pm e_\NS(x)$. 
We will show in the following subsection \ref{pentproof} that the  equations \eqref{superpentagon}
are equivalently to the integral identities
\begin{equation}\label{Ramanu}
\sum_{\sigma=0,1} \int_{-i\infty}^{i\infty} \frac{\ud\tau}{i} (-1)^{\rho_\beta\sigma}
e^{ \pi i \tau \beta} \frac{G_{\sigma+\rho_\alpha}(\tau + \alpha)}{G_{\sigma+1}
(\tau + Q)} = 2 \zeta_0^{-1} \frac{G_{\rho_\alpha}(\alpha)
G_{1+\rho_\beta}(\beta)}{G_{\rho_\alpha+\rho_\beta}(\alpha+\beta)} 
\end{equation}
which have been derived in  \cite{Hadasz:2007wi}.
 
 \subsection{The superpentagon equation}\label{pentproof}
 
 In this section we provide a proof of the superpentagon relations \eqref{superpentagon}
 for the functions $e_\NS(x)$ and $e_\R(x)$. We will show here that the equations \eqref{superpentagon} 
 are equivalent to the  analogs \eqref{Ramanu} of the Ramanujan summation formula. 
These formulae can be rewritten in terms of $e_\NS(x)$ and $e_\R(x)$
 as follows,
 \begin{subequations}\label{Ramanu-2}
 \begin{align*}
 \int \ud x e^{-\pi i x(u+c_\ub)} \left( \frac{e_\NS(x+c_\ub)}{e_\NS(x+v)} + \frac{e_\R(x+c_\ub)}{e_\R(x+v)} \right) &= 2 \chi_0 \frac{e_\NS(v + u + c_\ub)}{e_\NS(v) e_\NS(u)}, \\
 \int \ud x e^{-\pi i x(u+c_\ub)} \left( \frac{e_\NS(x+c_\ub)}{e_\NS(x+v)} - \frac{e_\R(x+c_\ub)}{e_\R(x+v)} \right) &= 2 \chi_0 \frac{e_\R(v + u + c_\ub)}{e_\NS(v) e_\R(u)}, \\
 \int \ud x e^{-\pi i x(u+c_\ub)} \left( \frac{e_\NS(x+c_\ub)}{e_\R(x+v)} + \frac{e_\R(x+c_\ub)}{e_\NS(x+v)} \right) &= 2 \chi_0 \frac{e_\R(v + u + c_\ub)}{e_\R(v) e_\NS(u)}, \\
 \int \ud x e^{-\pi i x(u+c_\ub)} \left( \frac{e_\NS(x+c_\ub)}{e_\R(x+v)} - \frac{e_\R(x+c_\ub)}{e_\NS(x+v)} \right) &= 2 \chi_0 \frac{e_\NS(v + u + c_\ub)}{e_\R(v) e_\R(u)},
\end{align*}
where $\chi_0 = e^{-i\pi(1- c_b^2)\slash 6} $. Taking the limit $v \to -\infty$ we can obtain the Fourier transforms
\begin{align*}
 \tilde f_+(u) &= \int \ud x e^{-\pi i x u} (e_\R(x)+e_\NS(x) ) = e^{-i\pi c_\ub u} \frac{2\chi_0}{e_\NS(u-c_\ub)} = \\
 &= \chi_0^{-1} e^{-i\pi u^2\slash2} e_\NS(c_\ub-u), \\
 \tilde f_-(u) &= \int \ud x e^{-\pi i x u} (e_\R(x)-e_\NS(x) ) = - e^{-i\pi c_\ub u} \frac{2\chi_0}{e_\R(u-c_\ub)} = \\
 &= i\chi_0^{-1} e^{-i\pi u^2\slash2} e_\R(c_\ub-u).
\end{align*}
\end{subequations}
Then, we can consider the matrix elements of the operators $f_r(\mathsf{X}) f_s(\mathsf{P}+\mathsf{X})$
between (generalised) eigenstates $\langle p |$ and $| p' \rangle$ of the operator $\mathsf{P}$ with
eigenvalues $p$ and $p'$, respectively:
\begin{equation*}
 \Xi_{rs} = \langle p | f_r(\mathsf{X}) f_s(\mathsf{P}+\mathsf{X}) | p' \rangle,
\end{equation*}
for $r,s=+,-$ and $[\mathsf{P},\mathsf{X}]=\frac{1}{i\pi}$. We have
\begin{align*}
 \langle p | f_r(\mathsf{X}) f_s(\mathsf{P}+\mathsf{X}) | p' \rangle &= \int \ud p'' \langle p | f_r(\mathsf{X}) | p'' \rangle\langle p''| f_s(\mathsf{P}+\mathsf{X}) | p' \rangle = \\
 & = \int \ud p'' e^{i\pi({p''}^2 - {p'}^2)\slash2} \tilde f_r(p-p'') \tilde f_s(p''-p'),
\end{align*}
where we used the identity between the matrix element of an arbitrary function $g$ and its Fourier transform $\tilde g$
\begin{equation*}
 \langle p| g(\mathsf{X}) | p' \rangle = \tilde g(p-p')\,,
\end{equation*}
and the fact that
\begin{equation*}
 g(\mathsf{X}+\mathsf{P}) = e^{\frac{i\pi}{2}\mathsf{P}^2} g(\mathsf{X}) e^{-\frac{i\pi}{2}\mathsf{P}^2} .
\end{equation*}
Let us consider in detail the case $r=+,s=+$. Then we can write, using \eqref{Ramanu-2},
\begin{align*}
 \Xi_{++} &= \int \ud p'' e^{\frac{i\pi}{2}({p''}^2 - {p'}^2)} \frac{e_\NS(p'-p''+c_\ub)}{e_\NS(p-p''-c_\ub)} e^{-\frac{i\pi}{2}(p''-p')^2}  e^{-i\pi c_\ub(p-p'')}= \\
 &= e^{-i\pi c_\ub(p-p')} \int \ud x e^{-i\pi x(p'+c_\ub)} \frac{e_\NS(x+c_\ub)}{e_\NS(x+p-p'-c_\ub)} = \\
 &= \chi_0 e^{-i\pi c_\ub(p-p')} \frac{1}{e_\NS(p-p'-c_\ub)} \left( \frac{e_\NS(p)}{e_\NS(p')} + \frac{e_\R(p)}{e_\R(p')} \right) .
\end{align*}
Therefore
\begin{equation*}
 f_+(\mathsf{X}) f_+(\mathsf{X}+\mathsf{P}) = e_\NS(\mathsf{P}) f_+(\mathsf{X}) e_\NS^{-1}(\mathsf{P}) + e_\R(\mathsf{P}) f_+(\mathsf{X}) e_\R^{-1}(\mathsf{P}) .
\end{equation*}
If one repeats the calculations for other possibilities, one finds
\begin{align*}
& f_-(\mathsf{X}) f_-(\mathsf{X}+\mathsf{P}) = -i( e_\NS(\mathsf{P}) f_+(\mathsf{X}) e_\NS^{-1}(\mathsf{P}) - e_\R(\mathsf{P}) f_+(\mathsf{X}) e_\R^{-1}(\mathsf{P}) ),\\
& f_+(\mathsf{X}) f_-(\mathsf{X}+\mathsf{P}) = -i( e_\R(\mathsf{P}) f_-(\mathsf{X}) e_\NS^{-1}(\mathsf{P}) - e_\NS(\mathsf{P}) f_-(\mathsf{X}) e_\R^{-1}(\mathsf{P}) ),\\
& f_-(\mathsf{X}) f_+(\mathsf{X}+\mathsf{P}) = e_\R(\mathsf{P}) f_-(\mathsf{X}) e_\NS^{-1}(\mathsf{P}) + e_\NS(\mathsf{P}) f_-(\mathsf{X}) e_\R^{-1}(\mathsf{P}) .
\end{align*}
Combining these relations one can easily obtain the system \eqref{oppenta} which was 
observed to imply the pentagon equation satisfied by $\mathsf{T}^{(1)}_{12}$.

\section{Superflips}\label{appendix2}

\begin{figure}[t]
\centering
\includegraphics[width=.975\textwidth]{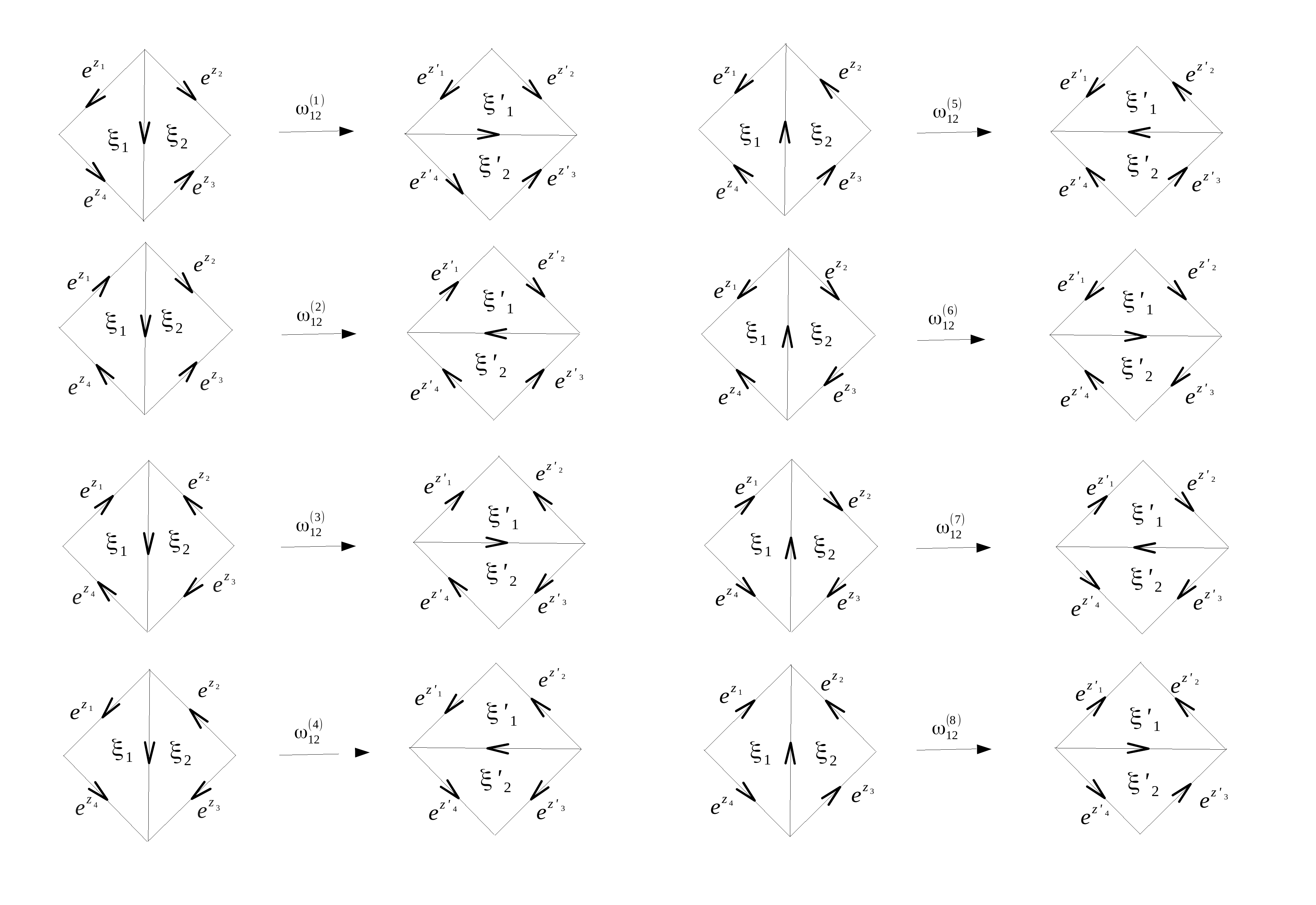}
\caption{Superflips for quadrilaterals without dots; cases 1-8.}
\label{superflips1}
\end{figure}

The superflip is a map which relates two different ways of triangulating a quadrilateral. In the case of super \TM theory, the triangles here should be interpreted as a dotted triangles, that is hexagons with Kasteleyn orientations. As we discussed in the main text, it is enough to consider flips between quadrilaterals with no dots, since one can remove dots by the action of push-outs. However, that still means that there are different ways of assigning Kasteleyn orientations to the long edges --- in fact, one has 8 possible ways to do that. In figure \ref{superflips1}  we present the full list of all of the possible superflips.


When considering Kashaev type coordinates it is necessary to use the decorated version of dotted triangulations. In the case of the quadrilaterals relevant for the flip map, decorated vertices should be chosen always as in figure \ref{flipkashaevclass}.




\section{Quantised flip $\mathsf{T}^{(1)}$}\label{appendix3}

In this section, we present the transformations of the quantised Fock coordinates under the flip that is given by the map $\mathsf{T}^{(1)}$. For the quadrilaterals on the figure \ref{flipkashaevclass}, the even Fock coordinates assigned to the edges are expressed as the operators on the $(L^2(\mathbb{R})\otimes\mathbb{C}^{1|1})^{\otimes2}$
\begin{align}
 \mathsf{Z}_e = e^{2 \pi b(\mathsf{q}_v - \mathsf{p}_v + \mathsf{p}_w)}\mathbb{I}_2, &\qquad \mathsf{Z}'_e = e^{2 \pi b(-\mathsf{q}_v + \mathsf{q}_w - \mathsf{p}_w)}\mathbb{I}_2, \\
 \mathsf{Z}_1 = e^{2 \pi b \mathsf{p}_v}\mathbb{I}_2, &\qquad \mathsf{Z}'_1 = e^{2 \pi b \mathsf{p}_v}\mathbb{I}_2, \\
 \mathsf{Z}_2 = e^{2 \pi b (\mathsf{q}_w - \mathsf{p}_w)}\mathbb{I}_2, &\qquad \mathsf{Z}'_2 = e^{2 \pi b(\mathsf{q}_v- \mathsf{p}_v)}\mathbb{I}_2, \\
 \mathsf{Z}_3 = e^{-2 \pi b \mathsf{q}_w}\mathbb{I}_2, &\qquad \mathsf{Z}'_3 = e^{-2 \pi b \mathsf{q}_w}\mathbb{I}_2, \\
 \mathsf{Z}_4 = e^{-2 \pi b \mathsf{q}_v}\mathbb{I}_2, &\qquad \mathsf{Z}'_4 = e^{2 \pi b \mathsf{p}_w}\mathbb{I}_2,
\end{align}
and the odd coordinates
\begin{align}
 \xi_1 = \sqrt{q^\frac{1}{2}-q^{-\frac{1}{2}}} \kappa\otimes\mathbb{I}_2, &\qquad \xi'_1 = \sqrt{q^\frac{1}{2}-q^{-\frac{1}{2}}} \kappa\otimes\mathbb{I}_2, \\
 \xi_2 = \sqrt{q^\frac{1}{2}-q^{-\frac{1}{2}}} \mathbb{I}_2\otimes\kappa, &\qquad \xi'_2 = \sqrt{q^\frac{1}{2}-q^{-\frac{1}{2}}} \mathbb{I}_2\otimes\kappa.
\end{align}
Those operators satisfy the algebraic relations as follows
\begin{align}
 [\mathsf{Z}_e, \mathsf{Z}_1] &= (1-q^{-4}) \mathsf{Z}_e \mathsf{Z}_1, \\
 [\mathsf{Z}_e, \mathsf{Z}_2] &= (1-q^{+4}) \mathsf{Z}_e \mathsf{Z}_2, \\
 [\mathsf{Z}_e, \mathsf{Z}_3] &= (1-q^{-4}) \mathsf{Z}_e \mathsf{Z}_3, \\
 [\mathsf{Z}_e, \mathsf{Z}_4] &= (1-q^{+4}) \mathsf{Z}_e \mathsf{Z}_4, \\
 [\mathsf{Z}_1, \mathsf{Z}_4] &= (1-q^{-4}) \mathsf{Z}_1 \mathsf{Z}_4, \\
 [\mathsf{Z}_2, \mathsf{Z}_3] &= (1+q^{+4}) \mathsf{Z}_2 \mathsf{Z}_3, \\
 [\mathsf{Z}_1, \mathsf{Z}_2] &= [\mathsf{Z}_1, \mathsf{Z}_3] = [\mathsf{Z}_2, \mathsf{Z}_4] = [\mathsf{Z}_3, \mathsf{Z}_4] = 0, \\
 [\mathsf{Z}_\alpha, \xi_i] &= 0, \\
 \{\xi_1,\xi_2\} &= 0, \\
 \{\xi_i,\xi_i\} &= 2\sqrt{q^\frac{1}{2}-q^{-\frac{1}{2}}} 1\otimes1.
\end{align}
Setting $q=e^{i\hbar\slash4}$ one can see that those commutation relations reproduce the classical Poisson bracket given by \eqref{superpoisson}.  

\noindent As an example, let us consider the transformation of the even variable $\mathsf{Z}'_1 = e^{2 \pi b \mathsf{z}'_1}$:
\begin{align*}
 &\mathsf{T}^{(1)-1}_{vw} \mathsf{Z}'_1 \mathsf{T}^{(1)}_{vw} = \\
 &=\frac{1}{4} e^{ \pi b\mathsf{p}_v} [(e_\NS^{-1}(u+ib)+e_\R^{-1}(u+ib))\mathbb{I}_2\otimes\mathbb{I}_2 - i (e_\R^{-1}(u+ib)-e_\NS^{-1}(u+ib))\kappa\otimes\kappa] \times \\
 &\times [(e_\NS(u-ib)+e_\R(u-ib))\mathbb{I}_2\otimes\mathbb{I}_2 - i (e_\R(u-ib)-e_\NS(u-ib))\kappa\otimes\kappa] e^{ \pi b\mathsf{p}_v} = \\
 &= \frac{1}{2} e^{ \pi b\mathsf{p}_v} \left\{ [e_\NS^{-1}(u+ib)e_\NS(u-ib) + e_\R^{-1}(u+ib)e_\R(u-ib)] \mathbb{I}_2\otimes\mathbb{I}_2 + \right. \\
 &\left. -i [e_\R^{-1}(u+ib)e_\R(u-ib) - e_\NS^{-1}(u+ib)e_\NS(u-ib)] \kappa\otimes\kappa \right\} e^{ \pi b\mathsf{p}_v} = \\
 &= e^{ \pi b\mathsf{p}_v} \left\{ [1 + e^{2\pi b (\mathsf{q}_v+\mathsf{p}_w-\mathsf{q}_w)} ] \mathbb{I}_2\otimes\mathbb{I}_2 + (q^{-\frac{1}{2}} - q^\frac{1}{2}) e^{\pi b  (\mathsf{q}_v+\mathsf{p}_w-\mathsf{q}_w)} \kappa\otimes\kappa \right\} e^{ \pi b\mathsf{p}_v} =\\
 &= \mathsf{Z}_1^\frac{1}{2} \left\{ (1 + \mathsf{Z}_e ) \mathbb{I}_2\otimes\mathbb{I}_2 + (q^{-\frac{1}{2}} - q^\frac{1}{2}) \mathsf{Z}_e^\frac{1}{2} \kappa\otimes\kappa \right\} \mathsf{Z}_1^\frac{1}{2} =\\
 &= \mathsf{Z}_1^\frac{1}{2} \left\{ (1 + \mathsf{Z}_e ) \mathbb{I}_2\otimes\mathbb{I}_2 - \mathsf{Z}_e^\frac{1}{2} \xione\xitwo \right\} \mathsf{Z}_1^\frac{1}{2} ,
\end{align*}
where we denoted $u= \mathsf{q}_v+\mathsf{p}_w-\mathsf{p}_v$ and used two times the shift relation of the quantum dilogarithm
\begin{align*}
 e_\R(x-ib) &= (1 - i(q^\frac{1}{2} - q^{-\frac{1}{2}}) e^{\pi bx} + e^{2\pi bx}) e_\R(x+ib), \\
 e_\NS(x-ib) &= (1 + i(q^\frac{1}{2} - q^{-\frac{1}{2}}) e^{\pi bx} + e^{2\pi bx}) e_\NS(x+ib).
\end{align*}
We can obtain the transformation property of the odd variable $\xi'_1$
\begin{align*}
 &\mathsf{T}^{(1)-1}_{vw} {\mathsf{Z}'}_1^{ \frac{1}{2}} \xi'_1 \mathsf{T}^{(1)}_{vw} = \sqrt{q^\frac{1}{2}-q^{-\frac{1}{2}}} \mathsf{T}^{(1)-1}_{vw} (e^{ \pi b \mathsf{p}_v} \kappa\otimes\mathbb{I}_2) \mathsf{T}^{(1)}_{vw} = \frac{1}{4} \sqrt{q^\frac{1}{2}-q^{-\frac{1}{2}}} e^{ \pi b \mathsf{p}_v}  \times \\
 &\times [(e_\NS^{-1}(u+ib)+e_\R^{-1}(u+ib))\mathbb{I}_2\otimes\mathbb{I}_2 - i (e_\R^{-1}(u+ib)-e_\NS^{-1}(u+ib))\kappa\otimes\kappa] \times \\
 &\times [(e_\NS(u)+e_\R(u))\mathbb{I}_2\otimes\mathbb{I}_2 - i (e_\R(u)-e_\NS(u))\kappa\otimes\kappa] \, \kappa\otimes\mathbb{I}_2 =\\
 &= \frac{1}{2} \sqrt{q^\frac{1}{2}-q^{-\frac{1}{2}}} e^{ \pi b \mathsf{p}_v} \left\{ [e_\NS^{-1}(u+ib)e_\R(u) + e_\R^{-1}(u+ib)e_\NS(u)] \mathbb{I}_2\otimes\mathbb{I}_2 + \right. \\
 &\left. -i [e_\R^{-1}(u+ib)e_\NS(u) - e_\NS^{-1}(u+ib)e_\R(u)] \kappa\otimes\kappa \right\} \, \kappa\otimes\mathbb{I}_2 =\\
 &= \sqrt{q^\frac{1}{2}-q^{-\frac{1}{2}}} e^{ \pi b \mathsf{p}_v} \left\{ \mathbb{I}_2\otimes\mathbb{I}_2 - q^\frac{1}{2} e^{\pi b  (\mathsf{q}_v+\mathsf{p}_w-\mathsf{p}_v)} \kappa\otimes\kappa \right\} \, \kappa\otimes\mathbb{I}_2 =\\
 &=  \mathsf{Z}_1^\frac{1}{2} (\xi_1 + q^\frac{1}{2} \mathsf{Z}_e^\frac{1}{2} \xi_2) = \mathsf{Z}_1^\frac{1}{4} (\xi_1 + \mathsf{Z}_e^\frac{1}{2} \xi_2) \mathsf{Z}_1^\frac{1}{4}.
\end{align*}
In this case we used the shift property of the quantum dilogarithm as well. In the analogous way, one can obtain the transformation properties of the rest of Fock variables in question.

\section{Super permutation}\label{appendixD}
 In this section we provide a computation of one of the super Ptolemy relations \eqref{rotationsandflips3} for $(j,k)=(4,1)$, which involves the operator $\Pi^{(1)}$ permuting our observables. Explicitly, we consider the relation
 \begin{equation}\label{susypermutationproof}
  \zeta_s \Pi^{(1)}_{(12)} = \mathsf{A}^{-1}_2 \mathsf{A}^{-1}_1 \mathsf{T}^{(4)}_{21} \mathsf{A}_1 \mathsf{T}^{(1)}_{12} .
 \end{equation}
The relation between two superflips is as follows
\begin{equation}
 \mathsf{T}^{(4)}_{12} = \mathsf{M}_1\mathsf{M}_2 \mathsf{T}^{(1)}_{12} \mathsf{M}_2. 
   \end{equation}
Let us denote $\alpha= \mathsf{q}_1 + \mathsf{p}_2 - \mathsf{q}_2$ and $\beta= \mathsf{q}_2 + \mathsf{p}_1 - \mathsf{q}_1$. Using that, the flips are expressed as
    \begin{align*}
 \mathsf{T}^{(1)}_{12} &= \frac{1}{2} 
 [(e_\R{(\alpha)}+e_{\NS}{(\alpha)}) \mathbb{I}_2\otimes\mathbb{I}_2 - i (e_\R{(\alpha)}-e_{\NS}(\alpha))\kappa\otimes\kappa ] e^{-\pi i \mathsf{p}_1\mathsf{q}_2}, \\
 \mathsf{T}^{(1)}_{21} &= \frac{1}{2} [(e_\R{(\beta)}+e_{\NS}{(\beta)}) \mathbb{I}_2\otimes\mathbb{I}_2 + i (e_\R{(\beta)}-e_{\NS}{(\beta)})\kappa\otimes\kappa ] e^{-\pi i \mathsf{p}_2\mathsf{q}_1}.
\end{align*}
In addition, lets recall that $\mathsf{A}$ acts on $\mathsf{p}$ and $\mathsf{q}$ as
\begin{align*}
 \mathsf{A}^{-1} \mathsf{q}\mathbb{I}_2 \mathsf{A} &= (\mathsf{p}-\mathsf{q})\mathbb{I}_2, \\ \mathsf{A}^{-1} \mathsf{p}\mathbb{I}_2 \mathsf{A} &= -\mathsf{q}\mathbb{I}_2.
\end{align*}
Using those formulae, we can evaluate the right hand side of \eqref{susypermutationproof}   
\begin{align*}
 \text{RHS} &= \frac{1}{4} \mathsf{A}_2^{-1} \mathsf{A}_1^{-1} \mathsf{M}_2\mathsf{M}_1 [(e_\R{(\alpha)}+e_{\NS}{(\alpha)}) \mathbb{I}_2\otimes\mathbb{I}_2 + i (e_\R{(\alpha)}-e_{\NS}(\alpha))\kappa\otimes\kappa ]\times\\ 
  &\quad \times \mathsf{M}_1 e^{-\pi i \mathsf{p}_1\mathsf{q}_2} 
  \mathsf{A}_1 [(e_\R{(\beta)}+e_{\NS}{(\beta)}) \mathbb{I}_2\otimes\mathbb{I}_2 - i (e_\R{(\beta)}-e_{\NS}{(\beta)})\kappa\otimes\kappa ] e^{-\pi i \mathsf{p}_1\mathsf{q}_2} =\\
 &= \frac{1}{4} \mathsf{A}_2^{-1} \mathsf{M}_2 [(e_{\R}(\mathsf{q}_2-\mathsf{p}_1)+e_{\NS}(\mathsf{q}_2-\mathsf{p}_1))\mathbb{I}_2\otimes\mathbb{I}_2 - i (e_{\R}(\mathsf{q}_2-\mathsf{p}_1)-e_{\NS}(\mathsf{q}_2-\mathsf{p}_1)) \kappa\otimes\kappa ] \times \\
 &\times [ (e_{\R}(\mathsf{p}_1-\mathsf{q}_2)+e_{\NS}(\mathsf{p}_1-\mathsf{q}_2) )\mathbb{I}_2\otimes\mathbb{I}_2 - i (e_{\R}(\mathsf{p}_1-\mathsf{q}_2)-e_{\NS}(\mathsf{p}_1-\mathsf{q}_2) )\kappa\otimes\kappa ]\\
 &\times e^{-\pi i \mathsf{p}_2(\mathsf{p}_1-\mathsf{q}_1)} e^{-\pi i \mathsf{p}_1\mathsf{q}_2} =\\
 &= \frac{1}{2} \mathsf{A}_2^{-1} \mathsf{M}_2 [ (e_\NS( \mathsf{q}_2-\mathsf{p}_1 ) e_\NS( -\mathsf{q}_2+\mathsf{p}_1 ) + e_\R( \mathsf{q}_2-\mathsf{p}_1 ) e_\R( -\mathsf{q}_2+\mathsf{p}_1 ) ) \mathbb{I}_2\otimes\mathbb{I}_2 + \\
 &- i (-e_\NS( \mathsf{q}_2-\mathsf{p}_1 ) e_\NS( -\mathsf{q}_2+\mathsf{p}_1 ) + e_\R( \mathsf{q}_2-\mathsf{p}_1 ) e_\R( -\mathsf{q}_2+\mathsf{p}_1 ) ) \kappa\otimes\kappa ] \times \\
 &\times e^{-\pi i \mathsf{p}_2(\mathsf{p}_1-\mathsf{q}_1)} e^{-\pi i \mathsf{p}_1\mathsf{q}_2} = \\
 &= \frac{1}{2} e^{i\pi c_\ub^2\slash2} e^{-\pi(1+2 c_\ub^2)\slash3} \mathsf{A}_2^{-1} \mathsf{M}_2 [ (e^{i\pi(-\mathsf{q}_2+\mathsf{p}_1)^2\slash2} + i e^{i\pi(-\mathsf{q}_2+\mathsf{p}_1)^2\slash2} ) \mathbb{I}_2\otimes\mathbb{I}_2 + \\
 &- i (- e^{i\pi(-\mathsf{q}_2+\mathsf{p}_1)^2\slash2} + i e^{i\pi(-\mathsf{q}_2+\mathsf{p}_1)^2\slash2} ) \kappa\otimes\kappa ] e^{-\pi i \mathsf{p}_2(\mathsf{p}_1-\mathsf{q}_1)} e^{-\pi i \mathsf{p}_1\mathsf{q}_2} = \\
 &= \frac{1+i}{2} e^{i\pi c_\ub^2\slash2} e^{-i\pi(1+2 c_\ub^2)\slash3}  \mathsf{M}_2 [  \mathbb{I}_2\otimes\mathbb{I}_2 + i \kappa\otimes\kappa ] \times \\
 &\times \underbrace{\mathsf{A}_2^{-1} e^{i\pi(-\mathsf{q}_2+\mathsf{p}_1)^2\slash2} e^{-\pi i \mathsf{p}_2(\mathsf{p}_1-\mathsf{q}_1)} e^{-\pi i \mathsf{p}_1\mathsf{q}_2}}_{e^{-i
 \pi/3} e^{i\pi/2} \mathsf{P}_{\rm b}} = \\
 &= e^\frac{i\pi}{4} e^{-i\pi(1+c_\ub^2)\slash6} \mathsf{M}_2 [ \mathbb{I}_2\otimes\mathbb{I}_2 + \kappa\otimes\kappa ] \mathsf{P}_{\rm b} =\zeta_s \mathsf{P}_{\rm f}\mathsf{P}_{\rm b} = \zeta_s \Pi_{12}^{(1)} = \text{LHS},
\end{align*}
which gives us the left hand side of the formula.



\end{document}